\documentclass[fleqn,usenatbib]{mnras}

\usepackage{newtxtext,newtxmath}
\usepackage[T1]{fontenc}
\usepackage{ae,aecompl}

\usepackage{bm}
\usepackage{subcaption}
\usepackage{graphicx}	% Including figure files
\usepackage{amsmath}	% Advanced maths commands
\usepackage{amssymb}	% Extra maths symbols

\graphicspath{{./Plots/}}
\title[Improving Catalogue Matching with Photometry]{Improving Catalogue Matching By Supplementing Astrometry with Additional Photometric Information}

\author[Tom J. Wilson and Tim Naylor]{
Tom J. Wilson,$^{1}$\thanks{E-mail: twilson@astro.ex.ac.uk}
and Tim Naylor$^{1}$
\\
$^{1}$School of Physics, University of Exeter, Stocker Road, Exeter EX4 4QL, UK\\
}

\date{Accepted 2017 October 12. Received 2017 October 12; in original form 2017 June 19}

\pubyear{2017}

\begin{document}
\label{firstpage}
\pagerange{\pageref{firstpage}--\pageref{lastpage}}
\maketitle

% Abstract of the paper
\begin{abstract}
The matching of sources between photometric catalogues can lead to cases where objects of differing brightness are incorrectly assumed to be detections of the same source. The rejection of unphysical matches can be achieved through the inclusion of information about the sources' magnitudes. The method described here uses the additional photometric information from both catalogues in the process of accepting or rejecting counterparts, providing approximately a factor 10 improvement in Bayes' factor with its inclusion. When folding in the photometric information we avoid using prior astrophysical knowledge. Additionally, the method allows for the possibility of no counterparts to sources as well as the possibility that sources overlap multiple potential counterparts. We formally describe the probability of two sources being the same astrometric object, allowing systematic effects of astrometric perturbation (by, e.g., contaminant objects) to be accounted for. 

We apply the method to two cases. First, we test IPHAS-\textit{Gaia} matches to compare the resulting matches in two catalogues of similar wavelength coverage but differing dynamical ranges. Second, we apply the method to matches between IPHAS and 2MASS and show that the method holds when considering two catalogues with approximately equal astrometric precision. We discuss the importance of including the magnitude information in each case. Additionally, we discuss extending the method to multiple catalogue matches through an iterative matching process. The method allows for the selection of high-quality matches by providing an overall probability for each pairing, giving the flexibility to choose stars known to be good matches.
\end{abstract}

% Select between one and six entries from the list of approved keywords.
% Don't make up new ones.

%%%%%%%%%%%%%%%%%%%%%%%%%%%%%%%%%%%%%%%%%%%%%%%%%%

%%%%%%%%%%%%%%%%% BODY OF PAPER %%%%%%%%%%%%%%%%%%

\section{Introduction}
\label{sec:cataloguematchintro}
The merging of two datasets, each containing a number of stars with photometric magnitudes, astrometric positions, and their related uncertainties is a fundamental process in many aspects of astrophysics. Broadband photometric measurements are crucial to gaining an understanding of a whole host of phenomena, from stellar physics to extragalactic luminosity functions. Frequently, a wide range of wavelengths will be required to compare theories to observations, and this is where it is important that the matches between different surveys are as accurate as possible.

The simplest matching methods only utilise the knowledge of the stars' positions and use a nearest-neighbour approach with a maximum cutoff distance when matching stars between two catalogues. Within the critical separation, two stars in two catalogues whose closest star in the other catalogue is each other will be assigned as a match, without consideration of either catalogue in a wider context, just considering each match on a pair-by-pair basis in isolation. We shall refer to this as proximity matching throughout this work. 

This crude catalogue matching process can be improved with the use of the astrometric information each detection provides. Sources are defined by their detected sky position, as well as a corresponding uncertainty in this measurement. It can be shown (e.g., Quetelet, as summarised by \citealp{Herschel:1857aa}) that the spatial probability distribution associated with this type of problem is described by a Gaussian. This leads to a better description of the pairing of sources between two catalogues, as it is then linked to the certainty to which the observations can be known, changing the effective matching radius.

However, as surveys probe increasingly fainter magnitudes, leading in turn to a correspondingly fainter saturation magnitude, the effects of matching two catalogues with significantly differing dynamical ranges is rapidly becoming an issue. If two cleaned catalogues were matched, one might contain a faint detection but have removed a bright object due to saturation effects, while the other might contain the bright object as a good detection but have the faint object below its sensitivity limit. If these two objects were within a given critical match separation, it could appear that two incompatible objects were nearest neighbours to one another, which would result in an unphysical object in the merged dataset.

To overcome incorrect matches, \citet{Sutherland:1992aa} defined the reliability of a source. They used knowledge of the source's ``type'' to identify optical counterparts to IRAS galaxies and overcome any faint object being assigned as a counterpart by nearest-neighbour matching. A thread in the literature (e.g., \citealp{Rutledge:2000aa}, \citealp{Fleuren:2012aa}) continues this method, supplementing astrometric knowledge with magnitude information available to create one-directional relationships between different types of object and their brightnesses. For example, \citet{Naylor:2013aa} map X-ray sources onto infrared (IR) detections, using the magnitudes in the IR catalogue but not those in the X-ray data.

\citet{Budavari:2008aa} symmetrised the procedure, considering magnitudes in both catalogues in question as equals to one another. However, they used astrophysical information to do so, fitting theoretical spectral energy distributions to each hypothetical match. This fitting then leads to a merged catalogue that is dependent on the assumptions made about the theoretical models.

Another line of work follows asymmetrical matching using solely the likelihood ratio of counterpart pairs (e.g., \citealp{Mann:1997aa}, \citealp{Brusa:2005aa}). The likelihood ratio between two stars from different catalogues is independent of the close presence of a second object in one of the catalogues. It does not consider the possibility of competition between objects in one catalogue for matches in the opposing catalogue. It therefore is a suboptimal solution in cases of high source density, where the chances of multiple sources being positionally close to a given object is high. In these cases the assumption that the distances between stars are significantly greater than the matching radius holds in neither catalogue. All competing hypotheses must therefore be considered jointly if any conclusion about the likelihood of an individual match is to be drawn, which may include the chance that multiple stars from either catalogue are potential matches to more than one star from the opposing catalogue. \citet{Naylor:2013aa} include the explicit probability of a non-pairing of the X-ray source to any of the IR detections when considering such asymmetric multiplicity.

While most of these methods focus on the matching of catalogues in the IR or X-ray wavelengths, there are examples of matching in other wavelengths in the literature. These include \citet{2017PASA...34....3L} in the radio and \citet{Pineau:2017aa} more generally across catalogues with relatively precise astrometry. However, in all cases the assumption that the astrometric probability is described by a Gaussian is still used. This does not correctly treat the effect systematic astrometric perturbations. These effects include proper motion and the contamination from faint stars (``crowding'', caused by the effects of finite pixel scale or PSF width). \citet{2017MNRAS.468.2517W} analyse these perturbations, discussing the relative effect they have on the matching separations.

These methods therefore do not simultaneously combine: 

\begin{itemize}
\item the creation of magnitude relationships between catalogues without the use of prior astrophysical knowledge;
\item photometric likelihoods which use these relationships bidirectionally, treating neither catalogue preferentially;
\item a symmetric process which allows for the matching of equal astrometric precision datasets;
\item the treatment of systematic effects in the astrometric detections of datasets;
\item the consideration of all positionally correlated detections simultaneously in the resulting match probabilities;
\item the explicit probability of a non-match of a star to any star in the opposing catalogue.
\end{itemize}

\begin{figure*}
    \centering
    \begin{subfigure}[b]{0.8\columnwidth}
        \centering
        \includegraphics[width=\columnwidth]{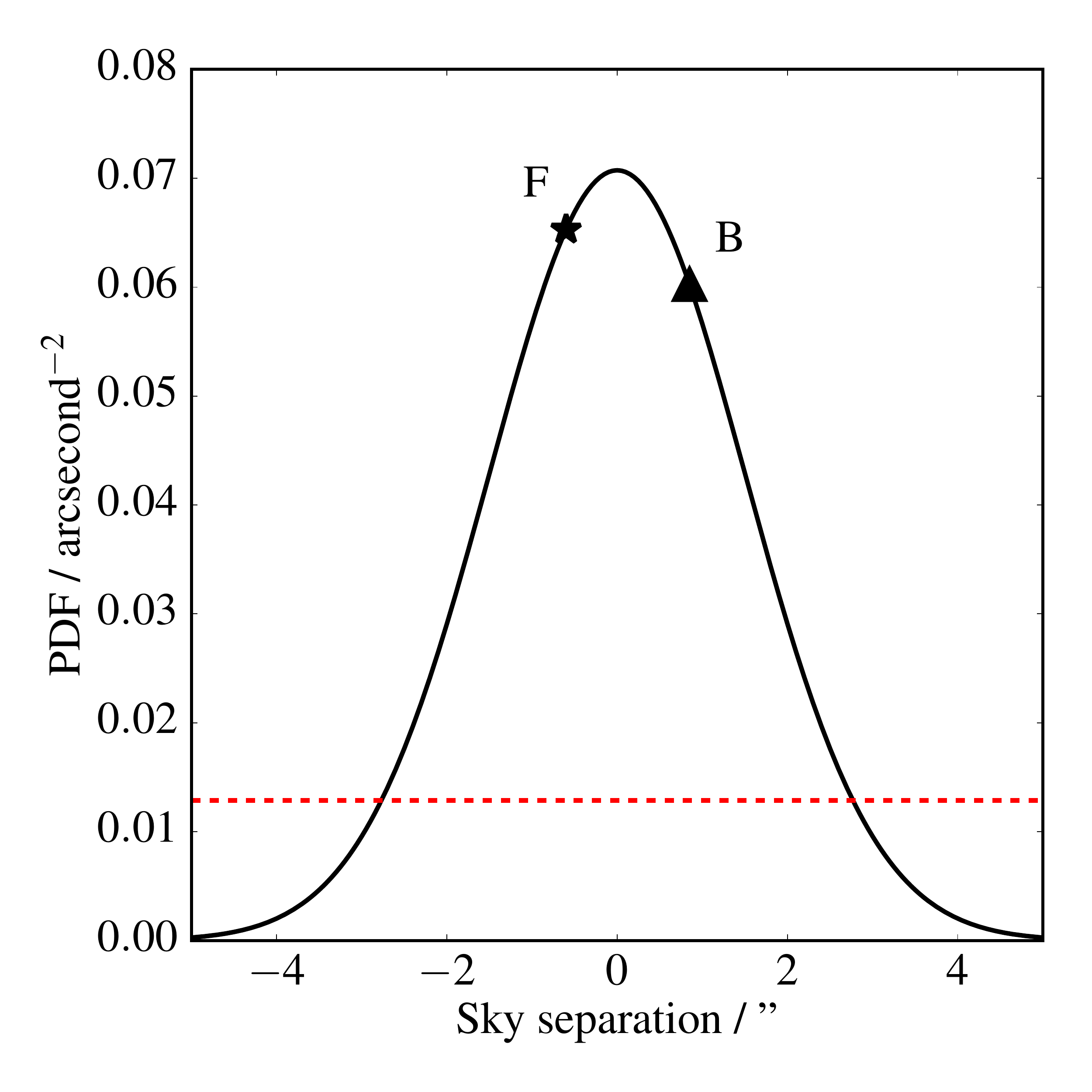}
        \caption{Schematic showing the probability of a detection of a star in one catalogue being a given distance from its detection in a second catalogue. The solid line shows the probability density of two stars being counterparts as a function of their radial offset. The dashed line shows the constant probability density of unrelated stars. Any stars at a smaller sky separation than the distance at which the two lines are of equal probability (i.e., where the line of counterpart probability is higher than the line denoting the density of unrelated stars) would be assigned as counterparts to one another in a matching scheme. Triangle and star markers denote the separations of the matches in hypotheses $B$ and $F$ respectively.}  
        \label{fig:poscomp}
    \end{subfigure}
    \hfill
    \begin{subfigure}[b]{1.2\columnwidth}  
        \centering 
        \includegraphics[width=\columnwidth]{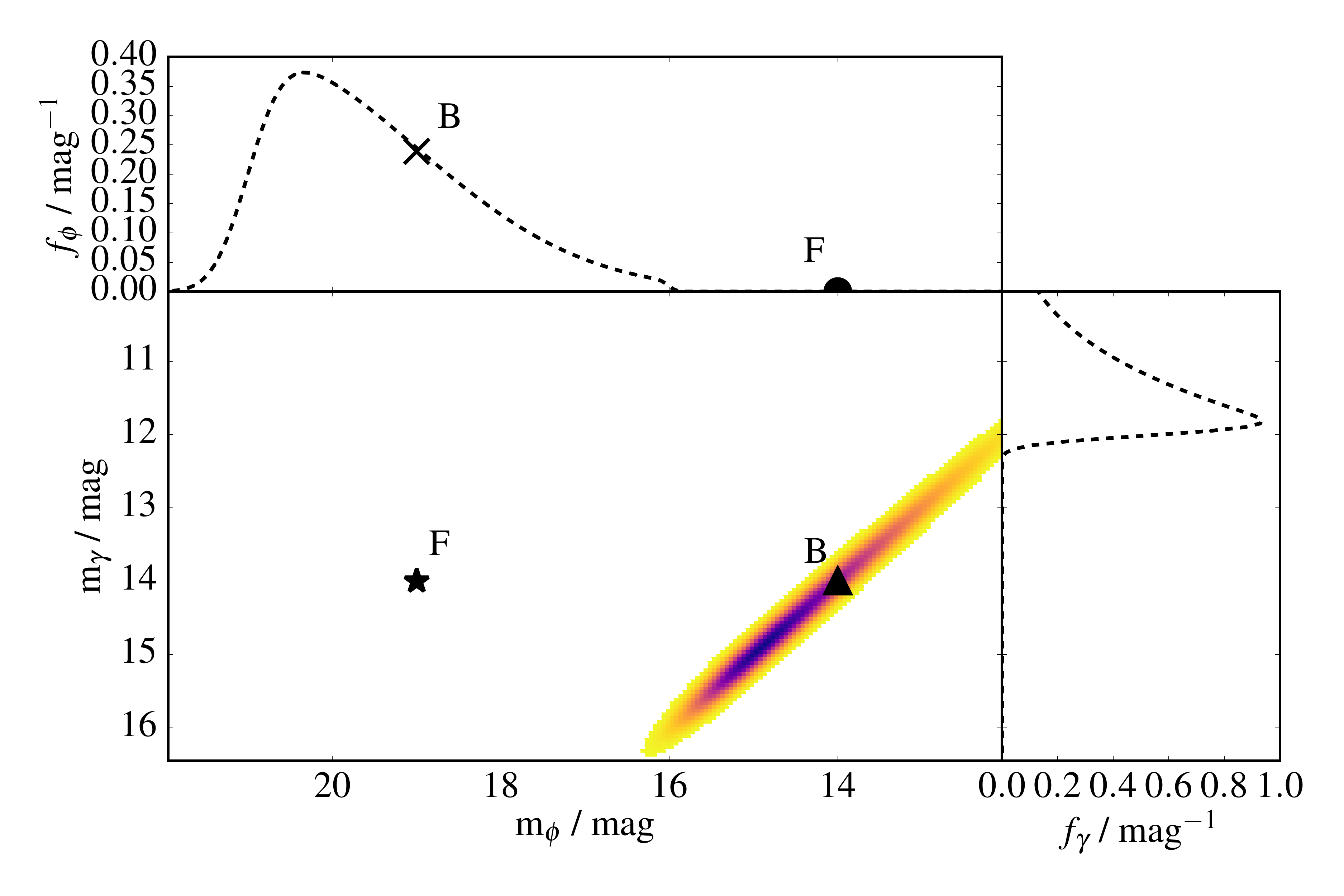}
        \caption{Figure showing the probability distribution for counterpart stars in two catalogues as a function of magnitude. Inset figures shows the distribution of unmatched star magnitude probability for the two catalogues. In this case the surveys used the same photometric filter and therefore have a high counterpart probability of their magnitudes matching. The probability of being an unmatched star is high in the case where a star in catalogue $\gamma$ is outside of the dynamical range of catalogue $\phi$, and vice versa. Also marked are the probability densities for two hypotheses. Hypothesis $B$ represents the case where two equal brightness ($m=14$) objects have been assigned as counterparts (triangle, main figure) while a faint object ($m=19$) in catalogue $\phi$ is unmatched (cross, inset figure). Hypothesis $F$ represents the alternative match case, where the bright object in catalogue $\phi$ is unmatched (circle, inset figure), and the faint catalogue $\phi$ object is matched to the object in catalogue $\gamma$ (star, main figure).}   
        \label{fig:magcomp}
    \end{subfigure}
    \caption{An example of star position and magnitude matching. Traditional matching would assign two detections as counterparts based purely on the positional probability, assigning the closest source only, preferring hypothesis $F$ on astrometric arguments alone. However, the addition of the magnitude information allows us to correctly match the true counterpart based on brightness, instead of simply positional correlation. The photometry allows for the pairing of the two objects with the magnitudes most likely drawn from an astrophysical object, accepting hypothesis $B$ with the inclusion of the extra parameter space.} 
    \label{fig:matchpositionexample}
\end{figure*}

Here we will derive a matching process that is fully symmetric between the catalogues being matched, generalising \citet{Naylor:2013aa}, highlighting the assumptions that any asymmetric matching processes implicitly require. We will also discuss how to extend the matching process to multiple catalogues simultaneously, and briefly touch upon a few ways to reduce the complexity of such a matching process. We begin by introducing the problem and giving an overview of how to overcome it in Section \ref{sec:matchqualitative}. Section \ref{sec:matchequation} gives a more rigorous derivation of the Bayesian formalism and the components of the equations. We then detail the forms that the astrometric, counterpart magnitude, and unmatched star magnitude distributions take, in Sections \ref{sec:astrofunctions} and \ref{sec:matchfunctions}. Section \ref{sec:catalogueapplication} gives two examples of the method applied to various catalogues. Section \ref{sec:multiplecatalogues} then describes how to extend the method to three or more catalogues, with concluding remarks in Section \ref{sec:conclusions}. Additionally, we demonstrate consistency with previous asymmetric matching methods by showing how the equations presented here reduce back to the one-directional forms given by \citet{Naylor:2013aa} in Appendix \ref{sec:matchreduction}. Table \ref{tab:symbols} defines symbols used throughout. 

\begin{table}
\centering
\begin{tabular}{c | l}
\hline
Symbol & Definition \\
\hline
$a$, $b$ & Semi-major and semi-minor star sky axes\\
$A_\phi$ & Counterpart PDF star area of consideration\\
$b_\phi$ & PDF of bright stars in $A_\phi$\\
$c(m_\gamma, m_\phi)$ & Symmetric counterpart magnitude PDF\\
$c(m_\gamma|m_\phi)$ & PDF of counterparts with magnitude $m_\phi$\\
$C(m_\gamma|m_\phi)$ & Integral of $c(m_\gamma|m_\phi)$ from $-\infty$ to $m_\gamma$\\
d$x$, d$y$ & Small sky widths defining sky cell area\\
d$m$ & Small range of stellar magnitudes\\
$f_\phi(m_\phi)$ & Unmatched catalogue $\phi$ star PDF\\
$F_\phi(m_\phi)$ & Integral of $f_\phi$ from $-\infty$ to $m_\phi$\\
$g(x_\gamma, x_\phi, y_\gamma, y_\phi)$ & PDF of two stars being counterparts given offset\\
$G(\Delta x, \Delta y)$ & PDF of two counterparts being offset in $x$ and $y$\\
$h_\phi$ & Astrometric uncertainty function of catalogue $\phi$\\
$i$, $j$, $k$, $l$ & Indices\\
$K$ & A normalisation\\
$m$ & The magnitude of a given star\\
$N_\mathrm{c}$ & Counterpart number density\\
$N_\phi$ & Unmatched catalogue $\phi$ number density\\
$n_\phi$ & Number of detected objects in catalogue $\phi$\\
$O$ & A normalisation\\
$p_\phi$ & PDF of all stars in catalogue $\phi$\\
$\mathcal{R}_Y$ & Radius defining circular Gaussian integral\\
$s$, $t$ & Indices\\
$T$ & Number of stars in a given magnitude range\\
$x$, $y$ & Cartesian coordinates\\
$Y$ & Fraction of Gaussian integral\\
$Z_{c\phi}$ & Fraction of stars with counterparts\\
$Z_\phi$ & Fraction of stars with at least one star inside $A_\phi$\\
$\alpha$, $\delta$ & Celestial Coordinates\\
$\gamma$ & A catalogue\\
$\epsilon$ & A catalogue\\
$\zeta$, $\lambda$ & Sets of catalogue detections\\
$\eta$ & Photometric likelihood ratio\\
$\theta$ & Position angle of sky axes\\
$\xi$ & Astrometric likelihood ratio\\
$\rho$ & Correlation of celestial sky axis uncertainties\\
$\sigma_\alpha$, $\sigma_\delta$ & Celestial sky axis uncertainties\\
$\phi$ & A catalogue\\
\hline
\end{tabular}
\caption{Table showing the definition of symbols used throughout.}
\label{tab:symbols}
\end{table}

\begin{table*}
\centering
\begin{tabular}{c | c}
\hline
Event & Notation\\
\hline
cell $i$ is empty in catalogue $\gamma$ & $E_\gamma^i$\\
cells $i$ and $j$ are occupied by a star that is in both catalogue $\gamma$ and $\phi$, respectively  & $S_{\gamma\phi}^{ij}$\\
cell $i$ is occupied by star in catalogue $\gamma$ that is not in catalogue $\phi$ & $U_\gamma^i$\\
\hline
\end{tabular}
\caption{Table showing the definitions of various events for catalogue matching.}
\label{tab:events}
\end{table*}

\section{Problem Setup}
\label{sec:matchqualitative}
Before we formalise the problem, it is useful to show qualitatively how the method works. For this purpose, we consider two catalogues that both contain detections in the same filter, with observations taken simultaneously with identical telescopes. One catalogue has good detections in the range $10 \leq m_\gamma \leq 16$, while the other catalogue has recorded sources with magnitudes $12 \leq m_\phi \leq 22$ There is a 100\% counterpart rate in the dynamical range of both catalogues, $12 \leq m \leq 16$. The smallest non-trivial problem of matching between the two catalogues is the case where one star in catalogue $\gamma$ and two stars in catalogue $\phi$ are positionally close to one another. All three stars are also sufficiently far away from all other stars that it can be assumed that no other star could be counterpart to any of the three of them. For illustration, let the given star in catalogue $\gamma$ have a magnitude $m_\gamma = 14$. The two stars in catalogue $\phi$ are one bright star, $m_\phi = 14$, the correct counterpart, and a faint star, $m_\phi = 19$, that is slightly closer to the star in catalogue $\gamma$ than its true counterpart. In our example, both stars in catalogue $\phi$ are close enough to be positionally likely to be matched with the star in catalogue $\gamma$. The two differing matches to the star in catalogue $\gamma$ are our hypotheses: $B$, in the case of the bright object match, and $F$, for the case where the faint object is the counterpart.

Figure \ref{fig:poscomp} shows an example schematic for the probability of two stars being matched given their sky separation. As the distance between their measured positions increases, the probability of the two stars being counterparts to one another decreases until they are more likely to be two unrelated stars. This is the point at which the counterpart probability density function (PDF) reaches the unmatched star probability density, indicated by the red dashed line. This probability is simply the chance of randomly placing unrelated stars in a small region of sky, based on the density of stellar sources nearby. If we were matching using this PDF alone, we would simply assign the stars as paired if their match probability is above the cut-off probability, or, equivalently, their separation is closer than the distance at which this transition occurs. In this case we would prefer hypothesis $F$, as the closest object to our star in catalogue $\gamma$ is the fainter of the two catalogue $\phi$ stars.

If we introduce the knowledge of the relationship between magnitudes in both catalogues, an example of which is shown in Figure \ref{fig:magcomp}, we now have a way to distinguish between our two sources in catalogue $\phi$. If we knew the intrinsic magnitude relationship between detections in each catalogue we could ask, based on the magnitude of two sources, whether they were likely to be the same star. In our example, both catalogues contain detections in the same filter, and therefore a detection in common between the two catalogues would measure the same brightness, to within experimental uncertainties. 

Shown as dashed lines in the insets to Figure \ref{fig:magcomp} are the probability densities of the objects in each catalogue ($\gamma$ or $\phi$) that do not have counterparts in the other catalogue ($\phi$ or $\gamma$). The unmatched PDF is the probability per unit magnitude that a star in catalogue $\gamma$, which does not have a corresponding entry in the catalogue $\phi$, is measured at its given brightness. These are those stars that are either too bright, having saturated in the survey images, or are too faint, having too low a signal-to-noise ratio to be counted as a good detection, to be recorded in catalogue $\phi$.

However, the probability of two stars being counterparts is a function of the brightness of both objects. This then leads to a two-dimensional function, an example of which is shown in the main panel of Figure \ref{fig:magcomp}. In our example, using the same filters means that our likelihood is effectively a straight line along $y = x$ in magnitude-magnitude space, albeit blurred by observational uncertainties. This is also a PDF, this time per square magnitude, of a star having detected magnitudes $m_\gamma$ and $m_\phi$ in the two catalogues respectively, given that it is the same object detected twice.

For these hypotheses it is expedient to consider some shorthand notation. We denote the astrometric probabilities of two stars being drawn from a distribution of counterparts given their separation as $g(m_*, m_1)$, and of a star not having a counterpart as $N$. The photometric probability of two stars having their quoted magnitudes given that they are counterparts is $c(m_*, m_1)$, and the probability of a star having its magnitude given that it is not related to the other catalogue is $f(m_1)$. We also define the star in catalogue $\gamma$ as $m_*$, the bright catalogue $\phi$ star as $m_1$, and the faint catalogue $\phi$ star as $m_2$.

Considering for the moment hypothesis $B$, we require a match between the star in catalogue $\gamma$ and the bright catalogue $\phi$ star, while also not matching the faint catalogue $\phi$ star. This we can write as

\begin{align}
\begin{split}
P(B|m_*, m_1, m_2) = \frac{g(m_*, m_1)c(m_*, m_1)N_\phi f_\phi(m_2)}{O},
\label{eq:hypoA}
\end{split}
\end{align}
where $O$ is a normalisation, which we will discuss below. Alternatively, we can consider the opposite match, 

\begin{align}
\begin{split}
P(F|m_*, m_1, m_2) = \frac{g(m_*, m_2)c(m_*, m_2)N_\phi f_\phi(m_1)}{O}.
\label{eq:hypoB}
\end{split}
\end{align}
We can also express the probability of the third case, in which neither star in catalogue $\phi$ is matched to the star in catalogue $\gamma$, as

\begin{align}
\begin{split}
P(C|m_*, m_1, m_2) = \frac{N_\gamma f_\gamma(m_*)N_\phi f_\phi(m_1)N_\phi f_\phi(m_2)}{O}.
\label{eq:hypoC}
\end{split}
\end{align}
In practice, we can dismiss this probability based on the assumption given previously that both catalogue $\phi$ stars are close enough to the catalogue $\gamma$ object to be considered likely. This means that $g(m_*, m_2) \gg N_\gamma N_\phi$. We include this third hypothesis for completeness. However, normalisation constant is simply the sum of the probability of all hypotheses, and thus

\begin{align}
\begin{split}
O = &N_\gamma f_\gamma(m_*)N_\phi f_\phi(m_1)N_\phi f_\phi(m_2) + \\&\,g(m_*, m_1)c(m_*, m_1)N_\phi f_\phi(m_2) + \\&\,g(m_*, m_2)c(m_*, m_2)N_\phi f_\phi(m_1).
\label{eq:hyponorm}
\end{split}
\end{align}

Considering our hypotheses $B$ and $F$, we only need focus on their photometric probabilities, as we have assumed that both stars in catalogue $\phi$ are at roughly equal sky separation from the catalogue $\gamma$ source, and thus $g(m_*, m_1)\simeq g(m_*, m_2)$.

Hypothesis $F$ (the faint star being the counterpart) leads to a low photometric probability density for all stars, with a low counterpart likelihood $c(m_*, m_2)$, and low field likelihood $f(m_1)$. However, the opposite hypothesis, $B$ (the bright star being the counterpart), has a high probability in both the photometric match between the two bright stars and the faint catalogue $\phi$ star being a field star. The main panel of Figure \ref{fig:magcomp} shows the probability densities for the counterpart matches for both hypotheses. Here the likelihood of the bright catalogue $\gamma$ object being the same object as the faint catalogue $\phi$ object photometrically is low, but the bright stars in both catalogues have a high probability of being the same source. Additionally, we further differentiate our hypotheses on the probability of the unmatched object. Along a similar line of reasoning, we can consider the unmatched object probability densities in the top inset figure. The rejected faint catalogue $\phi$ star in hypothesis $B$ has a high unmatched probability density, whereas hypothesis $F$ leads to a low unmatched star probability density.

We can use the combination of these two probability densities, for any matched and, just as usefully, unmatched objects, to help break any degeneracies in our astrometric matches. Such cases, where stars may have similar Mahalanobis distances, would be difficult to resolve with just the astrometric probability. This is especially significant when the astrometric probability is much higher than the unrelated source density against which a non-match is to compared. The result in our example is that while the bright catalogue $\phi$ object has a slightly larger sky separation (and would therefore not be matched astrometrically, by a nearest neighbour scheme or purely astrometric probability match; see Figure \ref{fig:poscomp} for comparison of the objects' sky separations), it is overwhelmingly more favourable as the counterpart. We can use the photometric information to correctly select the bright counterpart over the faint interloper.

While we have focused on the case where two stars are potential matches to a given object, we can also consider the trivial case. In this instance we have one star from each catalogue, and wish to determine whether they are counterparts or unrelated objects. If the stars were within the cut-off radius of a traditional proximity-matching method they would be paired automatically. However, the flexibility of the probability-based matching scheme allows us to directly compare the likelihood of the two stars being at their separations and magnitudes. We can examine both the case where they are the same star observed in two catalogues and the case where they are two different unrelated objects before considering them as counterparts.

\section{Constructing the Bayesian Framework}
\label{sec:matchequation}
Each photometric catalogue can be considered to be a three-dimensional position-position-magnitude cube. Each small square of sky plane is either filled with an object's detection, or blank and thus a non-detection. However, each position-position square that contains a star only has a filled cell at the recorded stellar magnitude. When matching two of these catalogues together, we are asking whether a given filled cell in catalogue $\gamma$ corresponds to a filled cell in catalogue $\phi$, or if they are unrelated.

Following a similar notation to that of section 2.1 of \citet{Sutherland:1992aa}, we define a ``cell'' to be have a volume $\mathrm{d}x\,\mathrm{d}y\,\mathrm{d}m$. We also define various events for detections and non-detections of objects in these cells, given in Table \ref{tab:events}. In terms of notation, for each event the subscript refers to the specific catalogue (in our case, either $\gamma$ or $\phi$), whereas the superscript refers to the individual cell (e.g., $i$ or $j$) in the given catalogue.

\subsection{The Match Hypotheses}
\label{sec:hypotheses}
Considering the case where one star in each catalogue is matched, and all other stars are unrelated, hypothesis $H_a$, we can write an expression for the likelihood of our data given this hypothesis,
\begin{align}
\begin{split}
P(D|H_a) \propto P&\left[S_{\gamma\phi}^{kl}\cap\left(\bigcap\limits_{i\neq k}U_\gamma^i\right)\cap\left(\bigcap\limits_{i{'}}E_\gamma^{i{'}}\right)\cap\left(\bigcap\limits_{j \neq l}U_\phi^j\right)\cap\left(\bigcap\limits_{j{'}}E_\phi^{j{'}}\right)\right].
\label{eq:onematch}
\end{split}
\end{align}
Here $S_{\gamma\phi}^{kl}$ is the probability that a given star occupies cell $k$ in catalogue $\gamma$ and cell $l$ in catalogue $\phi$, $E_\gamma^{i{'}}$ is the probability that cell $i{'}$ in catalogue $\gamma$ is empty, and $U_\gamma^i$ is the probability that cell $i$ is occupied by a star in catalogue $\gamma$ which is not in any cells in catalogue $\phi$. Equation \ref{eq:onematch} runs over $k$ and $l$, the cells containing only matched stars; $i$ and $j$, the cells filled with unrelated stars; and $i{'}$ and $j{'}$, the empty cells, for each catalogue respectively.

Now, if we consider the case where no stars are in common between the two catalogues, denoting it as $H_0$, we get a second hypothesis likelihood

\begin{align}
\begin{split}
P(D|H_0) \propto P\left[\left(\bigcap\limits_iU_\gamma^i\right)\cap\left(\bigcap\limits_{i{'}}E_\gamma^{i{'}}\right)\cap\left(\bigcap\limits_jU_\phi^j\right)\cap\left(\bigcap\limits_{j{'}}E_\phi^{j{'}}\right)\right],
\label{eq:nomatch}
\end{split}
\end{align}
where, again, $i$ and $j$ run over all filled cells and $i{'}$ and $j{'}$ run over all other cells.

At this point we can apply Bayes' rule to obtain hypothesis posteriors, given by

\begin{equation}
P(M|D) = \frac{P(D|M)P(M)}{P(D)}.
\label{eq:bayesrule}
\end{equation}
Here the evidence, $P(D)$, is simply the sum over all possible hypotheses; i.e., the sum over the null hypothesis $H_0$ and all possible combinations of $H_a$,

\begin{equation}
P(D) = P(D|H_0)P(H_0) + \sum\limits_aP(D|H_a)P(H_a).
\label{eq:bayesevidence}
\end{equation}
This requires a choice of prior. As we must accept any combination of unmatched and matched objects with equal probability, we have an indifferent prior, and thus $P(H_0) = P(H_a)$ for all $a$. We can then simply neglect it from the combination of equations \ref{eq:bayesrule} and \ref{eq:bayesevidence}. In addition, we can omit the sum over $i{'}$ and $j{'}$, as all empty cells remain empty in all hypotheses and are assumed to be independent of filled cells and each other. The terms simply cancel in the numerator and denominator of equation \ref{eq:bayesrule}. Thus our slightly modified version of equation 4 of \citet{Sutherland:1992aa} is 

\begin{align}
\begin{split}
&P(H_a|D) = \\
&\frac{P\left[S_{\gamma\phi}^{kl}\cap\left(\bigcap\limits_{i\neq k}U_\gamma^i\right)\cap\left(\bigcap\limits_{j \neq l}U_\phi^j\right)\right]}{P\left[\left(\bigcap\limits_iU_\gamma^i\right)\cap\left(\bigcap\limits_jU_\phi^j\right)\right]+\sum\limits_s\,\sum\limits_t\,P\left[S_{\gamma\phi}^{st}\cap\left(\bigcap\limits_{i\neq s}U_\gamma^i\right)\cap\left(\bigcap\limits_{j \neq t}U_\phi^j\right)\right]}.
\label{eq:rjold}
\end{split}
\end{align}

We can extend the independent cell assumption and split the probabilities. Therefore equation \ref{eq:rjold} becomes
\begin{align}
\begin{split}
&P(H_a|D) = \\
&\frac{P\left(S_{\gamma\phi}^{kl}\right)\prod\limits_{i \neq k}\,P\left(U_\gamma^i\right)\prod\limits_{j \neq l}P\left(U_\phi^j\right)}{\prod\limits_iP\left(U_\gamma^i\right)\prod\limits_jP\left(U_\phi^j\right) + \sum\limits_s\,\sum\limits_t\,P\left(S_{\gamma\phi}^{st}\right)\prod\limits_{i \neq s}\,P\left(U_\gamma^i\right)\prod\limits_{j \neq t}P\left(U_\phi^j\right)},
\label{eq:rjnew}
\end{split}
\end{align}
with the additional equation

\begin{align}
\begin{split}
&P(H_0|D) = \\
&\frac{\prod\limits_iP\left(U_\gamma^i\right)\prod\limits_jP\left(U_\phi^j\right)}{\prod\limits_iP\left(U_\gamma^i\right)\prod\limits_jP\left(U_\phi^j\right) + \sum\limits_s\,\sum\limits_t\,P\left(S_{\gamma\phi}^{st}\right)\prod\limits_{i \neq s}\,P\left(U_\gamma^i\right)\prod\limits_{j \neq t}P\left(U_\phi^j\right)}.
\label{eq:rjnew2}
\end{split}
\end{align}
Here $P(H_a|D)$ is a stand-in for $R_j$, the reliability of an object \citep{Sutherland:1992aa}, and we include the extra probability $P(H_0|D)$, introduced by \citet{Naylor:2013aa}. However, it is important to note that only \textit{unrelated} cells are independent, and we are therefore unable to separate the probabilities of a match between the two catalogues, and so must continue to consider $S_{\gamma\phi}^{st}$ jointly.

\subsection{Event Probabilities}
\label{sec:eventprobs}
We now require forms for each event, for which we follow the notation of \citet{Naylor:2013aa}. Position and magnitude are also assumed to be independent, and therefore are separable. For event $U$, the probability of an unrelated cell, we have 

\begin{equation}
P\left(U_\gamma^i\right) = N_\gamma\, \mathrm{d}x\, \mathrm{d}y\, f_\gamma(m_i)\,\mathrm{d}m,
\label{eq:uvfunction}
\end{equation}
where the probability of an unmatched star being in a given position is simply $N_\gamma$, the number density of unmatched stars, multiplied by $\mathrm{d}x \mathrm{d}y$, the cell sky area. Additionally, the probability of an unmatched star having magnitude $m$ to $m+\mathrm{d}m$ is $f_\gamma(m_i)$, the unmatched star magnitude distribution at $m_i$, multiplied by $\mathrm{d}m$.

The function for the probability of two stars matching between the two catalogues is slightly more involved. These require joint probabilities, which we write as

\begin{align}
\begin{split}
P\left(S_{\gamma\phi}^{kl}\right) = g(x_k, y_k, &x_l, y_l)\,\mathrm{d}x\, \mathrm{d}y\, \mathrm{d}x\, \mathrm{d}y\, c(m_k, m_l)\,\mathrm{d}m\,\mathrm{d}m
\label{eq:stfunction}
\end{split}
\end{align}
for now, and will expand each term separately. Here $g$ is the probability density, per degree$^4$, of two stars being counterparts to the same object with their recorded sky positions, while $c$ is the probability density, per square magnitude, that an object has its given quoted magnitudes in both catalogues.

No matter what combination of stars we have, we always consider the same volume $(\mathrm{d}x)^2(\mathrm{d}y)^2(\mathrm{d}m)^2$ for all stars. We therefore cancel the volume terms in equations \ref{eq:uvfunction} and \ref{eq:stfunction}, and make the change from pure probability to probability densities and a change from $P$ to $p$ in our notation.

\subsubsection{Astrometric Match Probability Density Function}
\label{sec:astropdfs}
The probability that the stars are counterparts requires the probability that star $k$ and $l$ are drawn from the same original sky position. This can be found by deriving the probability that the stars both originated from the same, but unknown, sky position $x_0$, $y_0$. It is relatively straightforward to compute the probability of two different detections of an object being at two sky positions given a known ``true'' position. However, it is more involved to obtain the probability of the two objects originating from the same position without prior knowledge. Handling this issue in a Bayesian fashion, we can marginalise over all ``true'' positions, giving us

\begin{align}
\begin{split}
&g(x_k, y_k, x_l, y_l) = \iint\limits_{-\infty}^{+\infty}\!p(x_k, y_k, x_l, y_l|x_0, y_0)p(x_0, y_0)\,\mathrm{d}x_0\,\mathrm{d}y_0 \\
&= \iint\limits_{-\infty}^{+\infty}\!h_\gamma(x_0 - x_k, y_0 - y_k) h_\phi(x_l - x_0, y_l - y_0)p(x_0, y_0)\,\mathrm{d}x_0\,\mathrm{d}y_0,
\label{eq:baymatch5a}
\end{split}
\end{align}
where $h_\gamma$ and $h_\phi$ are the rotationally symmetric (i.e., $f(x,\, y) = f(-x,\, -y)$) distributions of the astrometric uncertainties for catalogues $\gamma$ and $\phi$ respectively. We assign a flat prior on $x_0$ and $y_0$,

\begin{equation}
p(x_0, y_0) = N_\mathrm{c},
\label{eq:baymatch5b}
\end{equation}
the number of objects in common between the two catalogues per unit area. The details of how we calculate this number are described in Section \ref{sec:makecandfcomp}. Substituting equation \ref{eq:baymatch5b} into equation \ref{eq:baymatch5a} we obtain
\begin{align}
\begin{split}
g(x_k, y_k, x_l, y_l) = N_\mathrm{c}\iint\limits_{-\infty}^{+\infty}\!&h_\gamma(x_0 - x_k, y_0 - y_k)\times\\&h_\phi(x_l - x_0, y_l - y_0)\,\mathrm{d}x_0\,\mathrm{d}y_0.
\label{eq:baymatch5c}
\end{split}
\end{align}
We can substitute for the terms $\Delta x_{kl} = x_l - x_k$ and $\Delta y_{kl} = y_l - y_k$, giving

\begin{align}
\begin{split}
g(x_k, y_k, x_l, y_l) = N_\mathrm{c}\iint\limits_{-\infty}^{+\infty}\!&h_\gamma(x_0 - x_l + \Delta x_{kl}, y_0 - y_l + \Delta y_{kl})\times\\&h_\phi(x_l - x_0, y_l - y_0)\,\mathrm{d}x_0\,\mathrm{d}y_0.
\label{eq:baymatch5d}
\end{split}
\end{align}
Substituting $x = x_l - x_0$ and $y = y_l - y_0$ we obtain

\begin{align}
\begin{split}
g(x_k, y_k, x_l, y_l) &= N_\mathrm{c}\iint\limits_{-\infty}^{+\infty}\!h_\gamma(\Delta x_{kl} - x, \Delta y_{kl} - y)h_\phi(x, y)\,\mathrm{d}x\,\mathrm{d}y\\ &= N_\mathrm{c}\times(h_\gamma * h_\phi)(\Delta x_{kl}, \Delta y_{kl}).
\label{eq:convolveeq}
\end{split}
\end{align}
Here $(h_\gamma\, *\, h_\phi)(\Delta x_{kl}, \Delta y_{kl})$ denotes the convolution of the functions $h_\gamma$ and $h_\phi$, measured at position $\Delta x_{kl},\, \Delta y_{kl}$. To streamline our notation, we redefine equation \ref{eq:convolveeq} to be

\begin{equation}
g(x_k, y_k, x_l, y_l) = N_\mathrm{c}G(\Delta x_{kl}, \Delta y_{kl}).
\label{eq:convolveeq2}
\end{equation}

The resulting distribution is then a convolution of the two catalogues' individual astrometric uncertainty functions (AUFs; \citealp{2017MNRAS.468.2517W}), multipled by a prior term. This result is often quoted by other authors for the specific case where $G$ is Gaussian in both catalogues (e.g., equation 16 of \citealp{Budavari:2008aa}). In this simple case the convolution of the two functions is itself a Gaussian with uncertainty $\sigma_\mathrm{new}^2 = \sigma_k^2 + \sigma_l^2$, evaluated at $\Delta x_{kl},\, \Delta y_{kl}$. However, we know of no formal proof in the general case, although we note similarities between our equation \ref{eq:baymatch5a} and equation 9 of \citet{Budavari:2008aa} and, albeit without the prior term, equation 38 of \citet{Pineau:2017aa}. 

It should be noted that it cannot be assumed \textit{a priori} that $G$ will be a Gaussian, as the individual catalogue AUFs cannot themselves be assumed Gaussian. This is due to systematic effects such as proper motions, or the effects of faint contaminants within detected stars' point-spread functions (PSFs) on their measured positions \citep{2017MNRAS.468.2517W}. Our more general formalism allows for the inclusion of the treatment of such systematics (see Wilson \& Naylor, in prep. for a discussion of the effect this treatment has on the matching in highly contaminated crowded fields). Additionally, we note that this proof is only true for the specific case of matching two catalogues; see Section \ref{sec:multiplecatalogues} for the more general treatment of 3 or more catalogues.

\subsubsection{Photometric Match Probability Density Function}
\label{sec:photomatchpdf}
We also require the probability of two stars being related as a function of their respective magnitudes. If we had information about the intrinsic relationship between sources in both catalogues, we could marginalise over the stars' unknown ``true'' stellar magnitudes. This would be analogous to equation \ref{eq:baymatch5a}, and give

\begin{align}
\begin{split}
c(m_k, m_l) &= \iint\limits_{-\infty}^{+\infty}\!p(m_k, m_l|m_a, m_b)p(m_a, m_b)\,\mathrm{d}m_a\,\mathrm{d}m_b\\
&= \iint\limits_{-\infty}^{+\infty}\!p(m_k|m_a)p(m_l|m_b)p(m_a, m_b)\,\mathrm{d}m_a\,\mathrm{d}m_b.
\label{eq:baymatch5h}
\end{split}
\end{align}
The likelihoods in this case would be

\begin{align}
\begin{split}
p(m_k|m_a) = \frac{1}{\sqrt{2\pi}\sigma_k}\exp{\left(\frac{-(m_k - m_a)^2}{2\sigma_k^2}\right)}
\label{eq:baymatch5i}
\end{split}
\end{align}
and

\begin{align}
\begin{split}
p(m_l|m_b) = \frac{1}{\sqrt{2\pi}\sigma_l}\exp{\left(\frac{-(m_l - m_b)^2}{2\sigma_l^2}\right)},
\label{eq:baymatch5i2}
\end{split}
\end{align}
and $p(m_a, m_b)$ would represent the prior, intrinsic joint magnitude distribution on counterpart magnitudes $m_a$ and $m_b$. 

In practice, however, we cannot disentangle our observational uncertainties $\left(p(m_k|m_a),\ p(m_l|m_a)\right)$ and intrinsic relationships $\left(p(m_a, m_b)\right)$ from the data which measure $c(m_k, m_l)$, and we therefore measure $c$ directly. However, we include this description for symmetry and completeness.

\subsection{Combined Bayesian Probabilities}
\label{sec:combinedbayesprobs}

\subsubsection{One Match Equation Form}
\label{sec:onematchform}
For compact notation in this subsection, we define the following terms:

\begin{align}
\begin{split}
G(\Delta x_{kl}, \Delta y_{kl}) &= G^{kl}_{\gamma\phi}\\
c(m_k, m_l) &= c^{kl}_{\gamma\phi}\\
f_\gamma(m_i) &= f_\gamma^i.
\label{eq:compacteqs}
\end{split}
\end{align}
This notation follows a similar style to that previously, where each PDF ($G$, $c$, and $f$) has a subscript denoting which catalogue it refers to, and a superscript which identifies the star in the given catalogue. Our revised probabilities for $H_0$ and $H_a$ are therefore
\begin{align}
\begin{split}
&P(H_a|D) = \\
&\frac{N_\mathrm{c}G^{kl}_{\gamma\phi}c^{kl}_{\gamma\phi}\prod\limits_{i \neq k}\,N_\gamma f_\gamma^i\prod\limits_{j \neq l}N_\phi f_\phi^j}{\prod\limits_i\,N_\gamma f_\gamma^i\prod\limits_jN_\phi f_\phi^j + \sum\limits_s\,\sum\limits_t\,N_\mathrm{c}G^{st}_{\gamma\phi}c^{st}_{\gamma\phi}\prod\limits_{i \neq s}\,N_\gamma f_\gamma^i\prod\limits_{j \neq t}N_\phi f_\phi^j},
\label{eq:rjnew3}
\end{split}
\end{align}
and
\begin{align}
\begin{split}
&P(H_0|D) = \\
&\frac{\prod\limits_i\,N_\gamma f_\gamma^i\prod\limits_jN_\phi f_\phi^j}{\prod\limits_i\,N_\gamma f_\gamma^i\prod\limits_jN_\phi f_\phi^j + \sum\limits_s\,\sum\limits_t\,N_\mathrm{c}G^{st}_{\gamma\phi}c^{st}_{\gamma\phi}\prod\limits_{i \neq s}\,N_\gamma f_\gamma^i\prod\limits_{j \neq t}N_\phi f_\phi^j}.
\label{eq:rjnew4}
\end{split}
\end{align}
Equations \ref{eq:rjnew3} and \ref{eq:rjnew4} represent the fundamental result of this section, being generalised versions of previous formulations (e.g., equation 4 of \citealp{Sutherland:1992aa} or equation 7 of \citealp{Naylor:2013aa}). These equations give the probability of one star in catalogue $\gamma$ and one star in catalogue $\phi$ being counterparts, or the probability of there being no counterpart between stars in catalogues $\gamma$ and $\phi$, respectively. 

However, this formulation is limited, as shown by some simple example catalogues. Consider the case where both catalogue $\gamma$ and catalogue $\phi$ contain two objects each - $\gamma_{1, 2}$ and $\phi_{1, 2}$ respectively. The formulation used by \citet{Naylor:2013aa} assumes that each X-ray source (catalogue $\gamma$ object) does not compete with any other X-ray source for potential IR detection counterparts (catalogue $\phi$ objects). In such a case, equation 7 of \citet{Naylor:2013aa} would have two potential counterpart pairings, $\gamma_1\phi_1$ and $\gamma_1\phi_2$, as $\gamma_2$ is assumed to not be positionally close to these catalogue $\phi$ objects. In equation \ref{eq:rjnew3} we have lifted this assumption, allowing for two more hypotheses: $\gamma_2\phi_1$ and $\gamma_2\phi_2$, the pairing of our second catalogue $\gamma$ object with either catalogue $\phi$ object.

However, equation \ref{eq:rjnew3} assumes that, no matter how many stars are detected in either catalogue, at most one star was detected twice, and therefore only one star from one catalogue is a counterpart to one star in the other catalogue. This might be useful in many situations, where one catalogue is so sparse that two sources cannot possibly ``compete'' for the same source in the opposing catalogue (e.g., \citealp{Naylor:2013aa}), but is not necessarily the case in general. In crowded Galactic plane regions, for example, we may have a scenario where we cannot disentangle the recorded positions of multiple stars from each catalogue. We might reasonably assume that most of the objects recorded in both catalogues are the same objects detected twice. In this scenario, our example catalogues would have two additional hypotheses we must include: $\gamma_1\phi_1$ and $\gamma_2\phi_2$; and $\gamma_1\phi_2$ and $\gamma_2\phi_1$.

We can no longer make the assumption that we either have zero or one multiply detected object, as we have made throughout Section \ref{sec:matchequation} thus far. To account for the cases where we require the assigning of more than one counterpart pairing we must be able to express equations \ref{eq:rjnew3} and \ref{eq:rjnew4} in a more general form.

\subsubsection{Multiple Match Equation Form}
\label{sec:multiplematchform}
To account for multiple star pairings, equations \ref{eq:rjnew3} and \ref{eq:rjnew4} can be extended to any permutations of potential pairings between the catalogues $\gamma$ and $\phi$. For a given hypothesis, we wish to calculate the probability that there are $k$ matches between the two catalogues. Here $\zeta$ is a given $k$-permutation of catalogue $\gamma$, and $\lambda$ is a given $k$-combination of catalogue $\phi$. The use of permutations of one catalogue and combinations of the second catalogue avoids the repeated consideration of the same hypothesis - pairing $A$ with $B$ and $C$ with $D$ is the same as matching $C$ with $D$ and $A$ with $B$.

For example, if there are two matching stars between $\gamma$ and $\phi$ then $k = 2$. If there the four stars in $\gamma$, then we might have $\gamma = \{1, 2, 3, 4\}$. In this case, one potential subset of counterparts could be $\zeta = \{2, 4\}$. We require the probability that all stars which have been ``paired'' match, and all other stars are unmatched in both catalogues. $H_0$ is then the hypothesis that $k=0$, and $H_a$ is the hypothesis that there is one matched star in $\zeta$, paired with the star in $\lambda$. 

Our full equation is
\begin{align}
\begin{split}
P(\zeta&, \lambda, k|\gamma, \phi) = \frac{1}{K}\times\!\!\!\!\!\!\prod_{\delta\not\in\zeta\cap\delta\in\gamma}\!\!\!\!\!\!\!N_\gamma f_\gamma^\delta\!\!\!\!\!\!\prod_{\omega\not\in\lambda\cap\omega\in\phi}\!\!\!\!\!\!N_\phi f_\phi^\omega\prod\limits_{i=1}^kN_\mathrm{c}G^{\zeta_i\lambda_i}_{\gamma\phi}c^{\zeta_i\lambda_i}_{\gamma\phi},
\label{eq:baymatch11}
\end{split}
\end{align}
where $K$ is a normalisation constant, which we can generally express as the sum of the posterior probability of no matches plus the summation over all possible match number permutations. The normalisation requires a sum over three indices. First, the number of matches, $k$, from 0 to the number of objects in the smallest catalogue, resulting in a 100\% match rate, min($n_\gamma$, $n_\phi$). Second, each of the $k$-permutations of $\gamma$, the set of which we define as $\Gamma_k$. Finally, we must sum over each of the $k$-combinations of $\phi$, the set of which is $\Phi_k$. Thus
\begin{align}
\begin{split}
% K = \!\!\!\!\!\!\!\!\sum_{k=0}^{\min(n_\gamma, n_\phi)}\!\!\!\!\!\sum_{\zeta\in\Gamma_k}\sum_{\lambda\in\Phi_k} P(\zeta, \lambda, k|\gamma, \phi).
K = \!\!\!\!\!\!\!\!\sum_{k=0}^{\min(n_\gamma, n_\phi)}\!\!\!\!\!\sum_{\zeta\in\Gamma_k}\sum_{\lambda\in\Phi_k}\prod_{\delta\not\in\zeta\cap\delta\in\gamma}\!\!\!\!\!N_\gamma f_\gamma^\delta\!\!\!\!\!\!\prod_{\omega\not\in\lambda\cap\omega\in\phi}\!\!\!\!\!\!N_\phi f_\phi^\omega\prod\limits_{i=1}^kN_\mathrm{c}G^{\zeta_i\lambda_i}_{\gamma\phi}c^{\zeta_i\lambda_i}_{\gamma\phi}.
\label{eq:baymatch12}
\end{split}
\end{align}

While the equations presented are flexible in their application and set size, it is impractical to consider the entire dataset as one entity. We therefore limit our set size to those stars positionally close to another star in the set. This limitation results in a large number of star ``islands''. These islands could potentially reduce to the situation considered initially, with one star in one catalogue having multiple potential counterparts, for which equations \ref{eq:rjnew3} and \ref{eq:rjnew4} would be applicable. Typical number of stellar overlaps are $\leq$5, with the majority of stars only overlapped by 1-3 objects in the catalogue they are being matched to. We can therefore reduce the complexity in most cases back to that seen in equations \ref{eq:rjnew3} and \ref{eq:rjnew4}. In more complicated island permutations, with multiple stars in each catalogue under consideration, the more general equations \ref{eq:baymatch11} and \ref{eq:baymatch12} should be used.

In the next two sections we will expand our terms for $G$, $c$, and $f$, and detail how to calculate them.

\section{Functional Forms of Astrometric Distributions}
\label{sec:astrofunctions}
The astrometric PDF $G$ is defined for the two catalogue match as the convolution of the AUFs of the two stars in question (see Section \ref{sec:astropdfs}). As such, we require functions for the AUFs. For the rest of this paper we will assume that the probability of detecting a star with a given uncertainty, at a given offset ($x,\, y$) from its implied true origin, is given by a Gaussian. These AUFs describe how accurately the position of the star is known, which is vital for our probabilistic matching process. 

It can be shown \citep{2017MNRAS.468.2517W} that the empirical AUFs of a given catalogue may not be purely Gaussian, but are best described as broadened core distributions and large, non-Gaussian wings. These effects are caused by systematics such as proper motion or contamination from unresolved, faint objects inside the PSF of the bright star. However, for the purposes of testing our method in Section \ref{sec:catalogueapplication} we will focus on photometric catalogues with sufficiently small PSFs and number densities such that the average number of stars per PSF is low, which will limit the effect of the contamination to a few percent of stars at most.

In general, the AUFs can be two-dimensional elliptical Gaussians, meaning we require uncertainties in the orthogonal $\alpha$ (or right ascension), and $\delta$ (or declination) directions, as well as the correlation between the two, $\rho$. The transformations from semi-major axis $a$, semi-minor axis $b$, and position angle east of north $\theta$, if required, are given by

\begin{align}
\begin{split}
\sigma_\alpha &= \sqrt{a^2\sin^2(\theta) + b^2\cos^2(\theta)}\\
\sigma_\delta &= \sqrt{a^2\cos^2(\theta) + b^2\sin^2(\theta)}\\
\rho &= \frac{(a^2 - b^2)\sin(\theta)\cos(\theta)}{\sigma_\alpha\sigma_\delta}.
\label{eq:abt->xyrho}
\end{split}
\end{align}

For a two-dimensional PDF centered at the origin with covariance matrix

\begin{equation}
\bm{\Sigma} = \left( \begin{array}{cc} \sigma_\alpha^2 & \rho\sigma_\alpha\sigma_\delta \\ \rho\sigma_\alpha\sigma_\delta & \sigma_\delta^2 \end{array} \right).
\label{eq:covariancematrix}
\end{equation}
Our formulation of a given Gaussian AUF is then

\begin{align}
\begin{split}
h(\Delta \alpha, \Delta \delta, \sigma_\alpha, \sigma_\delta, \rho) &= \frac{\exp{\left(-\frac{1}{2\sqrt{1 - \rho^2}}\left(\frac{(\Delta \alpha)^2}{\sigma_\alpha^2} + \frac{(\Delta \delta)^2}{\sigma_\delta^2} - \frac{2\rho \Delta \alpha\Delta \delta}{\sigma_\alpha\sigma_\delta}\right)\right)}}{2\pi\sigma_\alpha\sigma_\delta\sqrt{1-\rho^2}}.
\label{eq:bivariatepdfs}
\end{split}
\end{align}
Note that when dealing with offsets in right ascension, we include the cosine of the declination to convert our separations to seconds of arc.

In our case we are considering the matching of three catalogues: IPHAS (\citealp{2005MNRAS.362..753D}; \citealp{Barentsen:2014tb}); 2MASS \citep{Skrutskie:2006um}; and \textit{Gaia} (\citealp{2016A&A...595A...1G}; \citealp{Gaia-Collaboration:2016aa}). For the rest of the paper we shall assume that the 2MASS and \textit{Gaia} astrometry are well modelled by Gaussians with uncertainties as quoted in their respective catalogues. IPHAS, however, does not quote individual source positional uncertainties, and we therefore use the relation given by \citet{King:1983aa},

\begin{equation}
\sigma_\alpha = \sigma_\delta = \sqrt{0.05"^2 + \left(\frac{\mathrm{FWHM_\mathrm{IPHAS}}}{2\sqrt{2 \log(2)} \times \mathrm{SNR_\mathrm{IPHAS}}}\right)^2}
\label{eq:king}
\end{equation}
where the full width at half maximum (FWHM) of the observational seeing is taken from the IPHAS catalogue for every star individually, and the signal-to-noise ratio (SNR) can be calculated from the statistical photometric uncertainty, also quoted individually for every star. The 0.05" is the typical systematic astrometric uncertainty. We use this combined uncertainty as our standard deviation in our Gaussian AUFs for the IPHAS data.

As we wish to calculate $G$, we must convolve our two Gaussian distributions together. To do so, we simply add the given covariance matrices of the two functions (equation \ref{eq:covariancematrix}) together, giving a new $\sigma_\alpha$, $\sigma_\delta$, and $\rho$, which we then use in equation \ref{eq:bivariatepdfs}.

\section{Functional Forms of Magnitude Distributions}
\label{sec:matchfunctions}
Now that we have the probability of correlation between two objects positionally, we must consider the probability of their relatedness in magnitude space. In this case, we must consider two possibilities. First, that each object in catalogue $\gamma$ is an unmatched object, unrelated to anything in catalogue $\phi$. Second, the two objects have magnitudes that have high likelihoods of being the same object detected in both catalogues. For these two cases we must build the counterpart probability density function, which we denote as $c(m_i, m_j)$, and the unmatched (``field''; \citealp{Naylor:2013aa}) star PDFs $f(m_i)$ and $f(m_j)$. $f$ is a PDF, the probability per unit magnitude of a star having its observed magnitude, given that it is unpaired (see, e.g., insets to Figure \ref{fig:magcomp}). $c$ is also a PDF, probability per unit $\gamma$ magnitude per unit $\phi$ magnitude, of two objects having their respective magnitudes given the assumption that they are counterparts to one another (Figure \ref{fig:magcomp}).

We construct these from the catalogues in situ. We must therefore consider the magnitudes of all stars in catalogue $\gamma$ positionally unrelated to any star in catalogue $\phi$ to build our unmatched magnitude distribution. Similarly we must consider the magnitudes of stars positionally close to one another to build our counterpart likelihood.

The unmatched star distributions are fairly straightforward, requiring merely the omission of any stars within sufficiently large circles of stars in the other catalogue, the details of which are described in Section \ref{sec:makecandfcomp}. We can then simply record the number of stars within each given narrow magnitude bin that remain to populate $f$. This will also remove some field stars, but under the assumption that the distribution of unrelated stars is positionally uncorrelated we still recover our distribution. 

\begin{figure}
    \centering
    \includegraphics[width=\columnwidth]{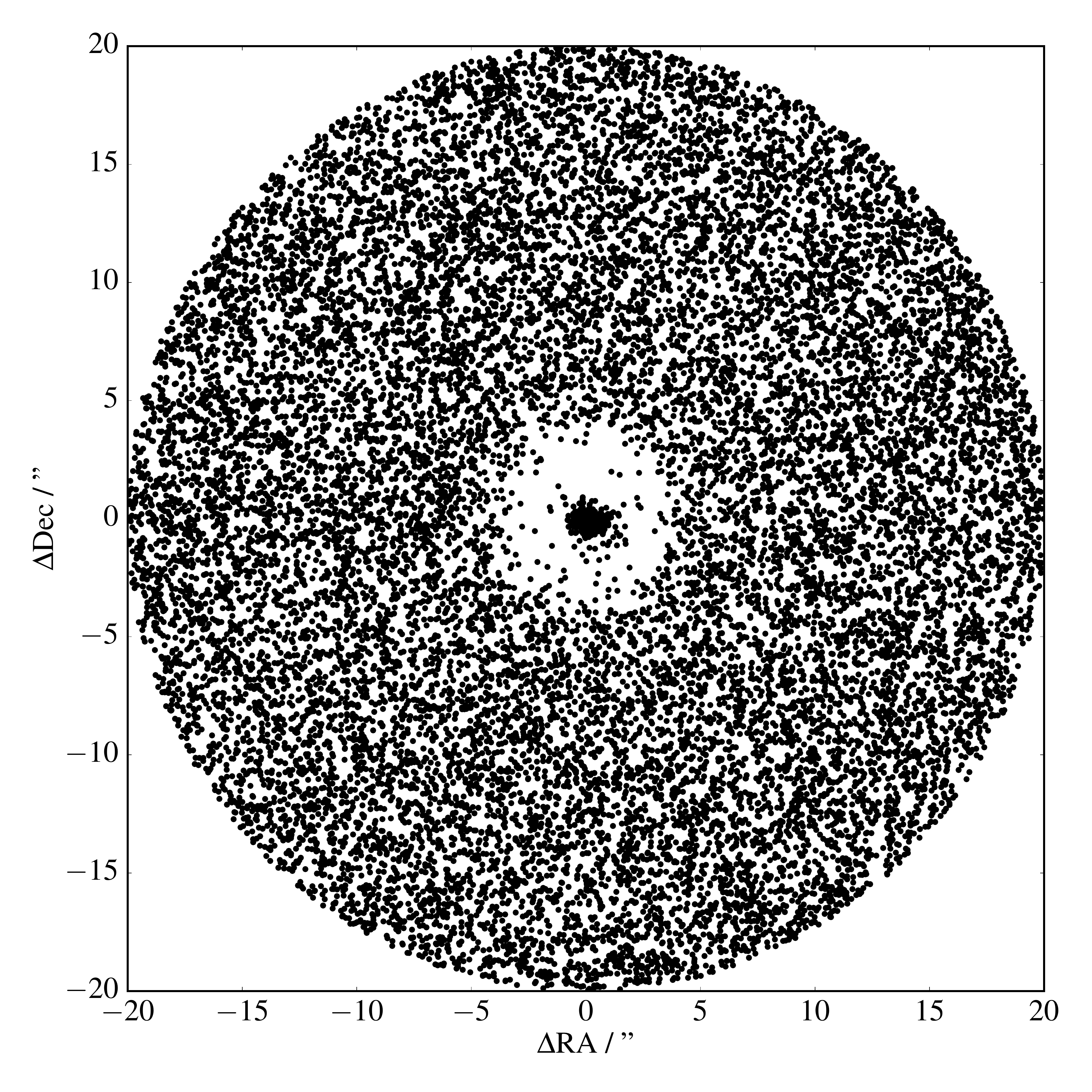}
    \caption{Figure showing the spatial separation of all 2MASS stars within 20" of \textit{Gaia} sources $15 \leq G \leq 15.25$, for a 5$^\circ \times 5^\circ$ slice of the Galactic plane. Background sources are seen at a constant density surrounding a clump of counterpart stars in the centre. However, the background density decreases within $\lesssim 3.5"$ due to the crowding out of the fainter background sources by bright counterparts.}
    \label{fig:crowding}
\end{figure}

Determining $c$ is rather more complex. Naively, one might simply record the magnitudes of those stars in catalogue $\phi$ close enough to the stars in question in catalogue $\gamma$ to be considered potential counterparts. However, there will be randomly placed unrelated stars that happen to lie close enough to another star to be considered a match, which will then be included in any distributions we create. To overcome this interloper problem, a sensible choice would then be to subtract a representative number of stars from each magnitude bin, using $f$ as the distribution to construct the ``background''. However, as shown in Figure \ref{fig:crowding} for the example of 2MASS sources positionally correlated with \textit{Gaia} sources $15 \leq G \leq 15.25$ at $120 \leq l \leq 125$, $0 \leq b \leq 5$, stars suffer from the crowding out of detections of stars fainter than themselves. We would therefore overestimate the number of faint field stars to be subtracted if we simply used the magnitudes of stars close to our chosen objects.

Instead of considering the closest stars to our sources, we can overcome the crowding effects by considering the brightest sources within a given radial offset, as developed in section 4 of \citet{Naylor:2013aa}. Using the bright star distribution, which is a density-independent measure, we can control for the decrease in the density of fainter objects. We can then correctly remove unrelated field objects from our distribution, obtaining a more robust counterpart distribution.

\begin{figure*}
    \centering
    \begin{subfigure}[b]{0.475\textwidth}
        \centering
        \includegraphics[width=\textwidth]{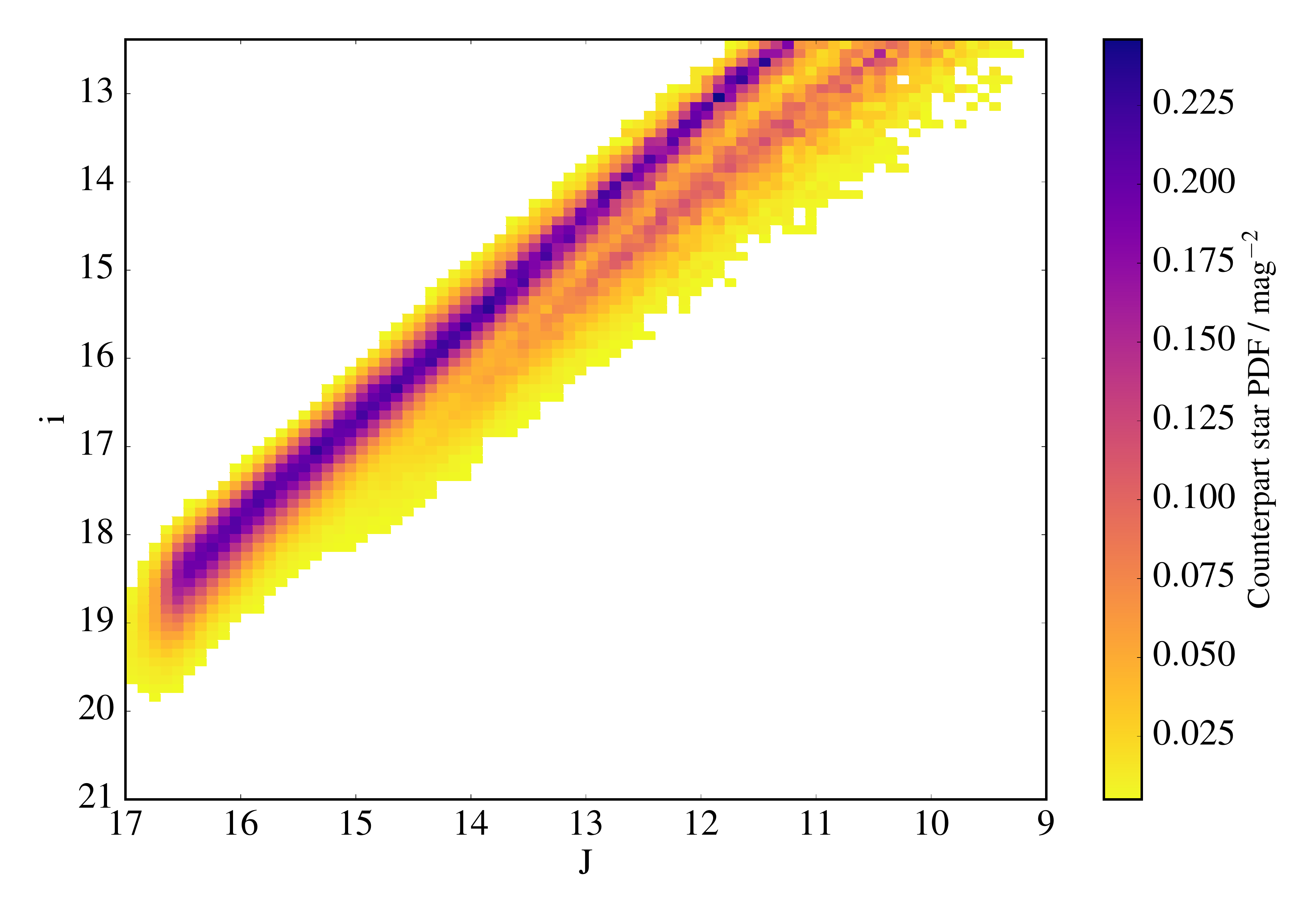}
        \caption{Un-corrected form of $c$, $c_\phi(m_\gamma|m_\phi)$, using IPHAS as the input catalogue.}  
        \label{fig:cmmcompa}
    \end{subfigure}
    % \hfill
    \begin{subfigure}[b]{0.475\textwidth}  
        \centering 
        \includegraphics[width=\textwidth]{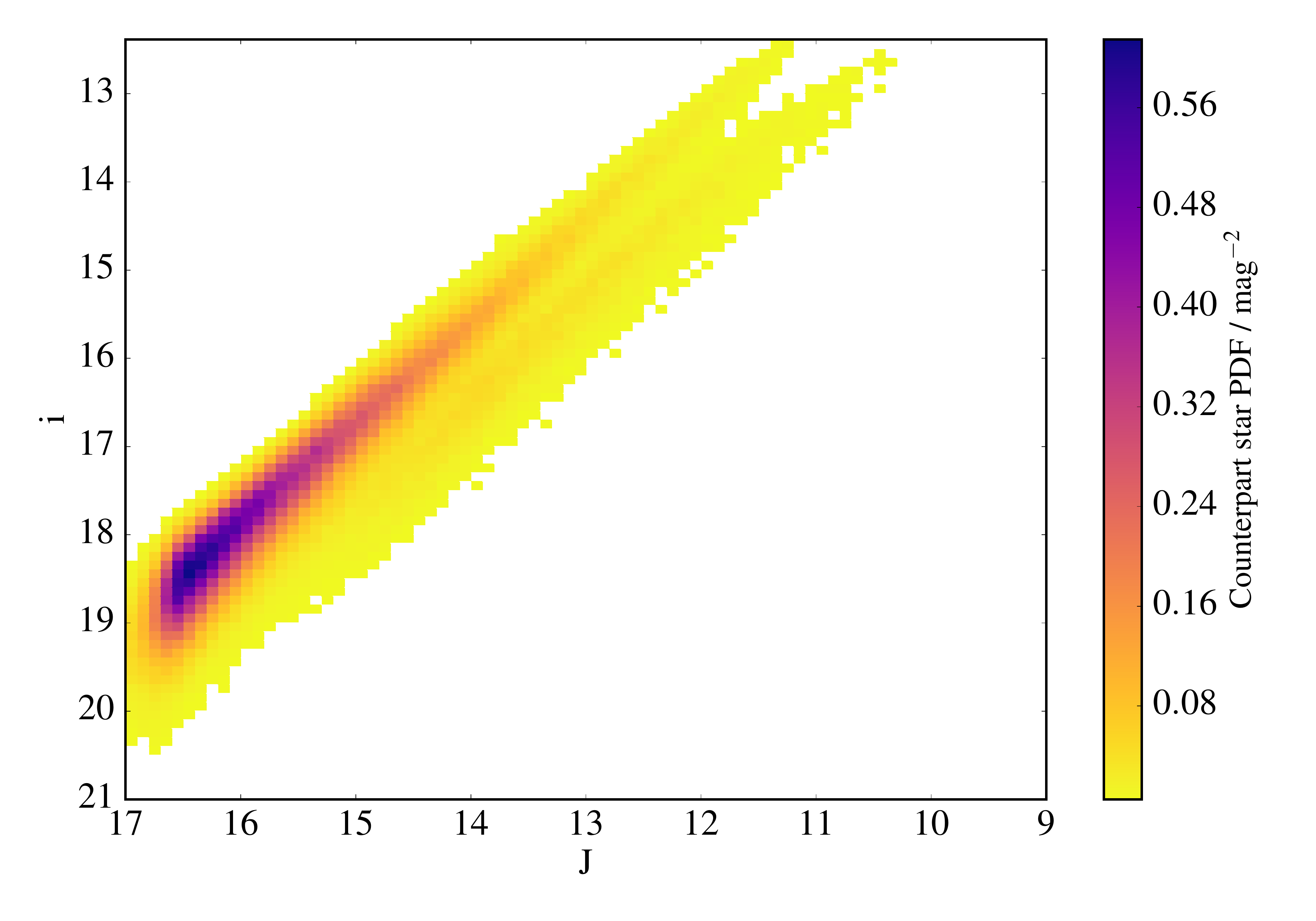}
        \caption{Corrected form of $c$, $c(m_\gamma, m_\phi)$, with IPHAS as the input.}   
        \label{fig:cmmcompc}
    \end{subfigure}
    % \vskip\baselineskip

    \begin{subfigure}[b]{0.475\textwidth}   
        \centering 
        \includegraphics[width=\textwidth]{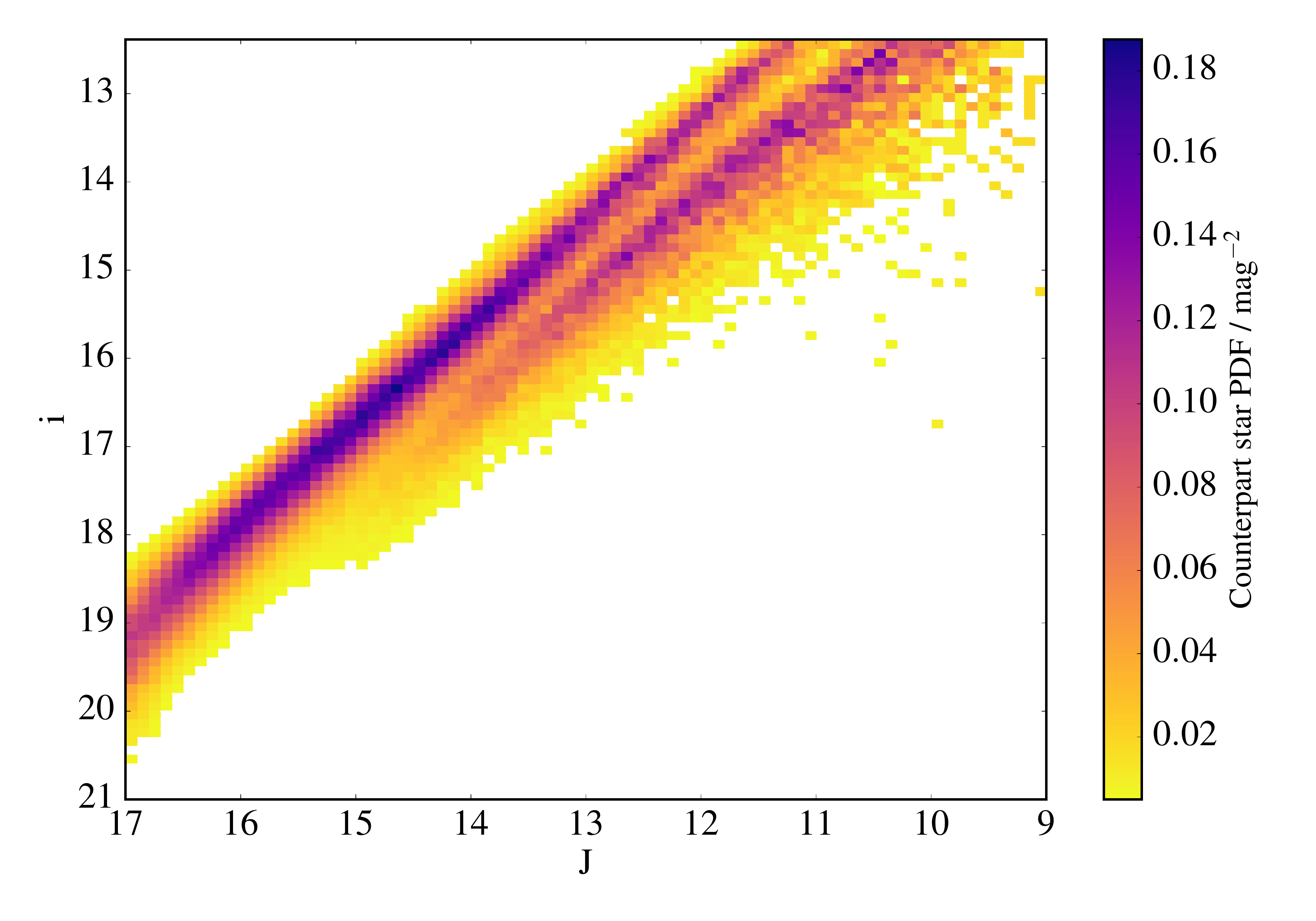}
        \caption{Un-corrected form of $c$, $c_\gamma(m_\phi|m_\gamma)$, using 2MASS as the input catalogue.}    
        \label{fig:cmmcompb}
    \end{subfigure}
    % \quad
    % \hfill
    \begin{subfigure}[b]{0.475\textwidth}   
        \centering 
        \includegraphics[width=\textwidth]{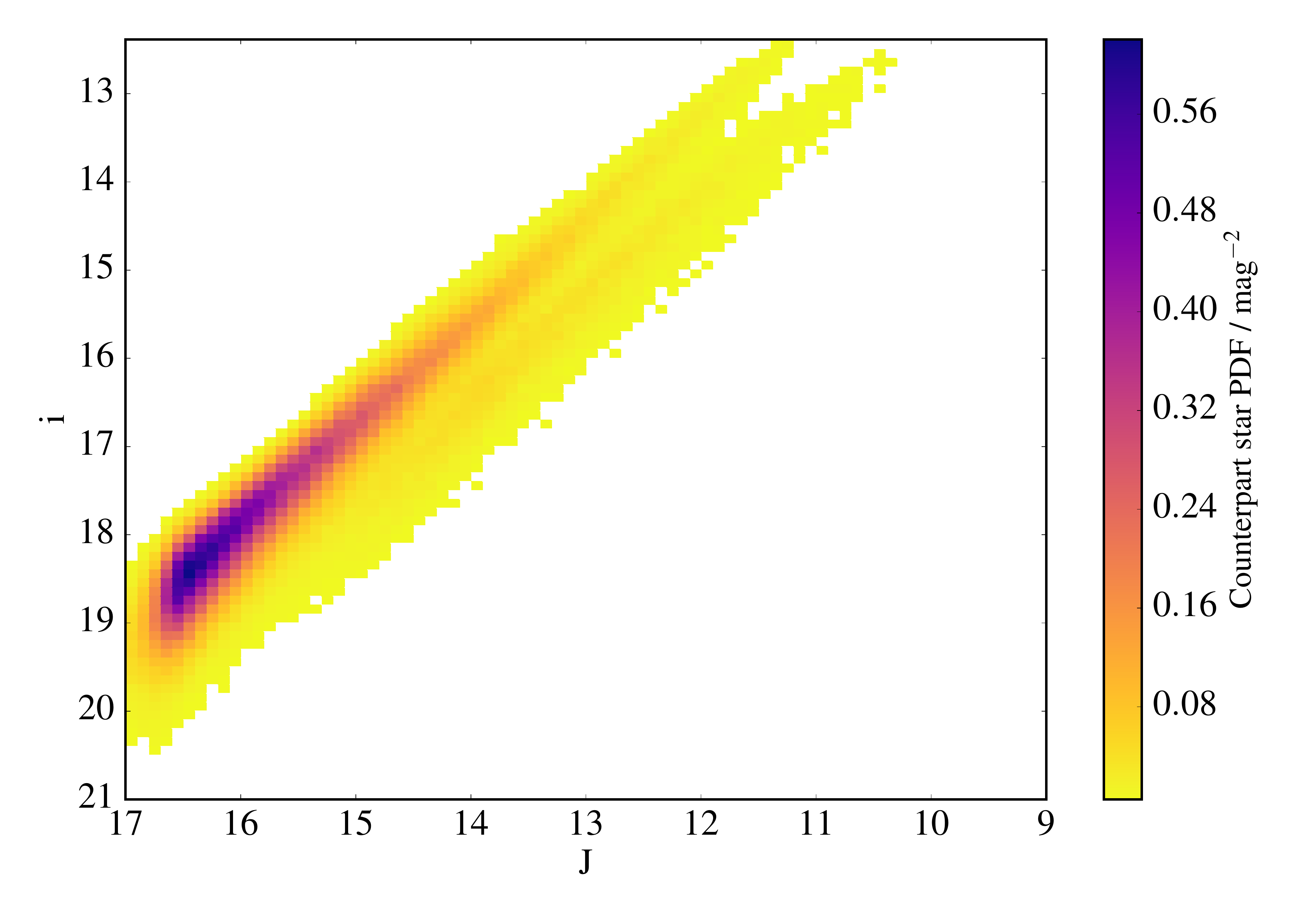}
        \caption{Corrected form of $c$, $c(m_\gamma, m_\phi)$, using 2MASS as the input catalogue.}  
        \label{fig:cmmcompd}
    \end{subfigure}
    \caption{Figure showing the effect asymmetry has on the overall counterpart probability density, for the comparison between the $J$ filter in 2MASS and the $i$ filter in IPHAS. Minimum colourmap is 0.005mag$^{-2}$ in all plots. If the symmetrisation step is not taken, the PDF only reflects one catalogue, leading to inconsistent results depending on which catalogue is used as the input. After symmetrisation, however, the PDFs are equivalent. Notation used assumes 2MASS as catalogue $\gamma$ and IPHAS as catalogue $\phi$, following discussion in Section \ref{sec:matchfunctions}.} 
    \label{fig:cmmcomparison}
\end{figure*}

However, \citet{Naylor:2013aa} only considered a one-sided problem, which effectively put the entirety of the second catalogue into one, very large, magnitude bin. The two-directional case requires the building of $c(m_\phi|m_\gamma)$ for each $m_\gamma$ to $m_\gamma+\mathrm{d}m$ bin, in turn. We therefore have, as our revised version of equation 16 of \citet{Naylor:2013aa}

\begin{align}
\begin{split}
Z_{c\gamma}\cdot c_\gamma(m_\phi|m_\gamma) = &Z_\gamma b_\gamma(m_\phi|m_\gamma)\exp{\left(A_\gamma N_\phi F_\phi(m_\phi)\right)} - \\&\left(1 - Z_{c\gamma} C_\gamma(m_\phi|m_\gamma)\right)A_\gamma N_\phi f_\phi(m_\phi).
\label{eq:zccstart}
\end{split}
\end{align}
Here $Z_{c\gamma}$ is the fraction of stars of magnitude $m_\gamma$ to $m_\gamma+\mathrm{d}m$ with counterparts inside a certain radial distance and $Z_\gamma$ is the fraction of stars of magnitude $m_\gamma$ to $m_\gamma+\mathrm{d}m$ with at least one star within the given radius. $b_\gamma(m_\phi|m_\gamma)$ is the distribution of the brightest stars within a radial offset of stars of magnitude $m_\gamma$ to $m_\gamma+\mathrm{d}m$. $A_\gamma$ is the average area inside the radial offsets for stars of magnitude $m_\gamma$ to $m_\gamma+\mathrm{d}m$ and $N_\phi$ is the number density of unmatched stars in catalogue $\phi$. $F_\phi(m_\phi)$ is the integral of the unmatched star distribution for catalogue $\phi$, $f_\phi(m_\phi)$, from $-\infty$ to $m_\phi$, and $C_\gamma(m_\phi|m_\gamma)$ is the integral of the counterpart star distribution, $c_\gamma(m_\phi|m_\gamma)$, from $-\infty$ to $m_\phi$.

There is an equivalent case with the switching of catalogues,

\begin{align}
\begin{split}
Z_{c\phi}\cdot c_\phi(m_\gamma|m_\phi) = &Z_\phi b_\phi(m_\gamma|m_\phi)\exp{\left(A_\phi N_\gamma F_\gamma(m_\gamma)\right)} - \\&\left(1 - Z_{c\phi} C_\phi(m_\gamma|m_\phi)\right)A_\phi N_\gamma f_\gamma(m_\gamma).
\label{eq:zccsymmetric}
\end{split}
\end{align}

These are not truly symmetric (see Figure \ref{fig:cmmcomparison} for comparison), because they are, effectively, expressions for $p(a|b)$ and $p(b|a)$; the conditional probabilities of $a$ given $b$ and of $b$ given $a$, respectively. However, we can easily obtain the joint probability of $a$ and $b$ by

\begin{equation}
p(ab) = p(a|b)p(b) = p(b|a)p(a).
\label{eq:cmmsymmetry} 
\end{equation}
The symmetrisation of $c$, from equations \ref{eq:zccstart} and \ref{eq:zccsymmetric}, is therefore

\begin{equation}
c(m_\gamma, m_\phi) = c_\gamma(m_\phi|m_\gamma) \cdot p_\gamma(m_\gamma) = c_\phi(m_\gamma|m_\phi) \cdot p_\phi(m_\phi).
\label{eq:cmmsymmetry2} 
\end{equation}
The effects of this additional probability are shown in Figure \ref{fig:cmmcomparison}, showing that our choice of input catalogue for construction of the magnitude-magnitude relationship does not affect the resulting PDF.

\section{Application to Photometry}
\label{sec:catalogueapplication}
To avoid using bad or unwanted data within individual surveys, we first clean the data using the criteria in Table \ref{tab:flagtable}. We have chosen three catalogues, \textit{Gaia}, 2MASS, and IPHAS, to highlight two important regimes for probabilistic matching. First, \textit{Gaia} and IPHAS are both optical surveys allowing for ease of comparison, but they have differing dynamical ranges, where IPHAS saturates at a fainter magnitude than \textit{Gaia} but also has a correspondingly fainter completeness limit. Second, the symmetrisation of the matching process means that we should be able to handle two catalogues with similar astrometric precision, which we test with an IPHAS-2MASS cross-match.

While the clean datasets ensure we do not include any spurious artifacts or other non-physical detections in our catalogues, we have also included some flags which remove true stellar detections. This means that our matches do not necessarily include every single source on the sky. Matching two cleaned datasets will result in some unpaired stars which, had we not removed poor detections, should have returned a corresponding detection in the opposing catalogue. This effect is similar to that discussed in Section \ref{sec:cataloguematchintro}, where the saturation of a star in one catalogue and the non-detection of a second star in the opposing catalogue can lead to a proximity mismatch of the two sources. 

One possible solution is to simply remove all stars in all catalogues surrounding a poor quality detection in any catalogue, at the cost of the removal of good quality data. This would allow for a more even matching, where all data were good quality in all potential matches. This, however, unneccessarily removes extra sources from our potential composite catalogue, and thus we choose to only remove the poor quality data. This has the additional advantage for this paper of leaving these ``orphan'' stars in our catalogues, which provide a good test of the rejection of star pairings based on their photometry. We will see later in this section that we successfully return these stars as unmatched field objects.

More generally this effect is seen in crowded fields, where one catalogue, with high angular resolution, is matched to another, less able to resolve individual sources. This results in the effect, also discussed later in this section, where the bright resolved object is matched to the single contaminated source in the opposing catalogue. We then return the faint source in the high resolution catalogue as an unmatched object. Care must therefore be taken when matching two catalogues of differing resolution to not misinterpret these as stars with corresponding missing detections below the sensitivity of the survey in question. The ``completeness limit'' of a survey, often quoted as a single magnitude, is therefore highly dependent on the interplay of the resolving power of the survey and the local density of sources.

\begin{table*}
\centering
\begin{tabular}{c | c | c}
\hline
Catalogue & Flag & Criteria\\
\hline
\textit{Gaia} & Non-stellar & astrometric\_excess\_noise > 0.865mas and astrometric\_excess\_noise\_sig > 2\\
 & Low Quality & astrometric\_excess\_noise > 0.865mas and astrometric\_excess\_noise\_sig $\leq$ 2; or\\
& & astrometric\_n\_good\_obs\_al + astrometric\_n\_good\_obs\_al < 60; or matched\_observations $\leq$ 8\\
\hline
2MASS & Non-stellar & ``Galcontam'' or ``Mpflag'' flags set\\
 & Outside Dynamic Range & ``Blend'' flag == 0; or ``Read'' flag == 0 or 3; or Mag == NaN; or $\sigma_{\mathrm{Mag}}$ == NaN\\
 & Low Quality & ``Photqual'' flag is not ``A'', ``B'', or ``C''; or ``Read'' flag is not 1 or 2; or \\
 &  & ``Blend'' flag is not 1, 2, 3; or ``Contam'' flag is not ``0'' or ``c''\\
\hline
IPHAS & Non-stellar & $p_{\mathrm{star}} < 0.9$\\
 & Outside Dynamic Range & Mag == NaN; ``Saturated'' flag set; or $\sigma_{\mathrm{Mag}}$ == NaN\\
 & Low Quality & ``Deblend'' or ``BrightNeighbour'' flagged; $\sigma_{\mathrm{Mag}} > 0.1$; or \\
 & & $\lvert\mathrm{Mag}-\mathrm{AperMag1}\rvert > 3\sqrt{\sigma_\mathrm{Mag}^2 + \sigma_{\mathrm{AperMag1}}^2} + 0.03$\\
\hline
\end{tabular}
\caption{Table showing the various flags for non-stellarity, detection and photometric quality for the catalogues used. In cases where flags refer to a specific filter, \textit{Gaia} only uses the $G$ filter, IPHAS uses the $r$ and $i$ filters, while 2MASS is cleaned using the $J$, $H$, and $K_\mathrm{s}$ filters.}
\label{tab:flagtable}
\end{table*}

\subsection{Reducing Computational Complexity}
\label{sec:reducecompcomplex}
Equation \ref{eq:baymatch12} is too computationally expensive to treat the entirety of a catalogue as one set, as discussed in Section \ref{sec:multiplematchform}. We reduce the complexity by initially assuming that there is no overlap between stars drawn from the same catalogue, which we shall refer to as ``internal independence''. However, we must account for the chance of a star from catalogue $\phi$ being positionally close to two stars from catalogue $\gamma$, even if those original stars are not positionally overlapping one another. Such ``external dependencies'' would stop us being able to treat stars in catalogue $\gamma$ independently and force us to consider them as part of a larger set. This assumption is borne out in the one-directional case considered by \citet{Naylor:2013aa}, in which they were able to assume their X-ray dataset was internally independent, but, due to the multiplicity of the potential matches, the IR data were not independent of one another. Here we are simply generalising this to both catalogues, creating ``groups'' of both sets of, e.g., X-ray and IR, detections. We have therefore relaxed the assumption that internal independency holds for one of the catalogues, but must break our matches up into groupings which have inter-group independency, for computational purposes.

Throughout the next two sections we discuss certain ``radial'' distances, which we define formally here for clarity and notation succinctness. These radial distances, $\mathcal{R}_Y$, are defined as the distance at which a certain percentage ($Y$) of a circular integral of a two-dimensional Gaussian is enclosed. They are the solution to the equality

\begin{equation}
\iint\limits_{x{'}^2 + y{'}^2 \leq \mathcal{R}_Y^2}\! G(x{'}, y{'})\mathrm{d}x{'} \mathrm{d}y{'} = Y,
\label{eq:raddist}
\end{equation}
where $G$ is the convolution of two sources' AUFs (see Section \ref{sec:astropdfs} for definition and discussion).

To break our matches into independent groupings we first iterate over the entirety of one catalogue, assigning as potential counterparts to each star those stars in the other catalogue which appear within a certain ``merging radius''. These potential counterpart lists are merged in cases, as previously, where two stars could potentially match to the same star in the opposing catalogue. These mergers give a complete list of ``islands'' which are independent of each other but must be considered jointly within. We are extremely conservative with our rejecting of potential counterparts, using a large merging radius.

To calculate the radius at which we consider objects close enough to be related, we first find the star at the 95$^\mathrm{th}$ percentile uncertainty ellipse area - $\pi ab$ - for each catalogue. This gives uncertainties that avoid significant outliers, but that are larger than those of the vast majority of the survey. The semi-major and semi-minor axes of those stars are then used to construct $G$. We define stars to be positionally close to one another if they are separated by less than $\mathcal{R}_{0.997}$ ($\simeq3.4\sigma$ for a circular, two-dimensional Gaussian), our critical merging radius.

Each island is then fed into equations \ref{eq:baymatch11} and \ref{eq:baymatch12}, and the most probable arrangement is accepted, with stars being assigned as counterparts or unmatched stars. We can then either accept this permutation or reject it as uncertain depending on whether its probability lies above a certain threshold. For example, we can accept the most likely permutation, no matter the probability; accept permutations with $p > 0.5$, where the highest probability permutation outweighs all other permutations; or we can be more strict, requiring $p > 0.8$ (e.g., \citealp{2013ApJS..209...32B}). The probabilities in this section are accepted where the overall permutation probability $p > 0.5$; i.e., where the most likely permutation is more likely than all other options combined.

\subsection{Constructing \textit{f} and \textit{c} computationally}
\label{sec:makecandfcomp}
To calculate $f$, we must ``cut out'' a large section around each catalogue $\gamma$ star in catalogue $\phi$, to avoid any possibility of introducing the true counterpart to our unmatched probabilities. However, due to the large variations in precision for detections, we must consider each star individually when avoiding potential counterparts. When masking a given star in catalogue $\gamma$, we ignore any stars in catalogue $\phi$ within a certain distance. This distance is found by finding the star in catalogue $\phi$ in the same ``island'' as the catalogue $\gamma$ star in question with the largest astrometric uncertainties. We then use the two stars' AUFs to create a new $G$ distribution, and find $\mathcal{R}_{0.9}$. It is this radius inside which catalogue $\phi$ objects close to the catalogue $\gamma$ star are ignored. $Y = 0.9$ was chosen as a tradeoff between two requirements. First, we wish to minimise the contamination from counterparts appearing in our uncorrelated sample, nominally at the 10\% level but mitigated by the fact that $G$ always uses the largest possible uncertainties. Second, if possible we should mitigate against low number statistics, avoiding overly large ``cut out'' radii caused by the integration of $G$ to large distances. In addition to calculating $f_\gamma$ and $f_\phi$, we calculate $N_\gamma$ and $N_\phi$ from the area the catalogue covers after the star masks were applied, subtracting the total area masked by the calculated radial offsets. 

To construct $c$, we use equation \ref{eq:zccstart}, and therefore require the building of distributions of $b$, the bright star distribution. For this, we define radii for each star in a given catalogue in a similar way to when we construct $f$, except we use $\mathcal{R}_{0.63}$, the 0.6$\times$FWHM optimal result from \citet{Naylor:1998aa}. This radius trades off between minimising the effects of unmatched stars in our distributions while ensuring we still have enough counterparts to ensure good number statistics. $N_\mathrm{c}$ was calculated by integrating each $Z_{c\phi}\cdot c_\phi(m_\gamma|m_\phi)$ to obtain $Z_{c\phi}$, because each $c_\phi$ slice should be normalised if our $b_\phi$ slice and $f_\gamma$ are normalised. This then gives us, for the magnitude slice, the fraction of stars with counterparts within $\mathcal{R}_{0.63}$. To obtain the overall fraction of stars with counterparts, we must divide by the fraction expected, $Y = 0.63$. Once we have the fraction of input objects which have counterparts, we can obtain the number density of counterparts by multiplying by the number density of sources in the small magnitude slice. Repeating this for all magnitudes, we sum the density of counterparts for each input magnitude slice to obtain the total counterpart number density, $N_\mathrm{c}$.

Throughout this section we will be comparing number densities of matches, for both the matched counterparts and unrelated field stars. For the one dimensional density these are simply the number of objects with a magnitude $m_\gamma$ to $m_\gamma + \Delta m_\gamma$, $T$, divided by bin width $\Delta m_\gamma$. In the two dimensional case the number density is the number of objects with magnitude $m_\gamma$ to $m_\gamma + \Delta m_\gamma$ and magnitude $m_\phi$ to $m_\phi + \Delta m_\phi$, $T$, divided by bin widths $\Delta m_\gamma\Delta m_\phi$. We will consider three sources of counts: the probability-based counterpart matches ($T_\mathrm{prob}$), the proximity-based matches ($T_\mathrm{prox}$), and the probability-based unmatched objects. These number densities, while not normalised, are comparable to the PDFs $c$ and $f$. The number density of counterparts is related to $AN_cc$, where $A$ is the area of sky under consideration, while $AN_\phi f_\phi$ is the equivalent field star number density.

\subsection{Probabilistic Matches}
\label{sec:probmatches}
We have constructed both our astrometric uncertainty functions and our counterpart and unmatched star magnitude PDFs, and so can begin to match our catalogues. For the test cases, the two catalogues were extracted for a 25 square-degree area of the sky, $120^\circ \leq l \leq 125^\circ$, $0^\circ \leq b \leq 5^\circ$, and any stars which did not contain at least one filter flagged as a detection (either good or low quality) were discarded. Then $c$ and $f$ were constructed for each filter - $i$ for IPHAS, $J$ for 2MASS, and $G$ for \textit{Gaia} - along with the corresponding number densities.

\subsubsection{IPHAS vs \textit{Gaia}}
\label{sec:iphasvsgaia}
\begin{figure*}
    \centering
    \includegraphics[width=\textwidth]{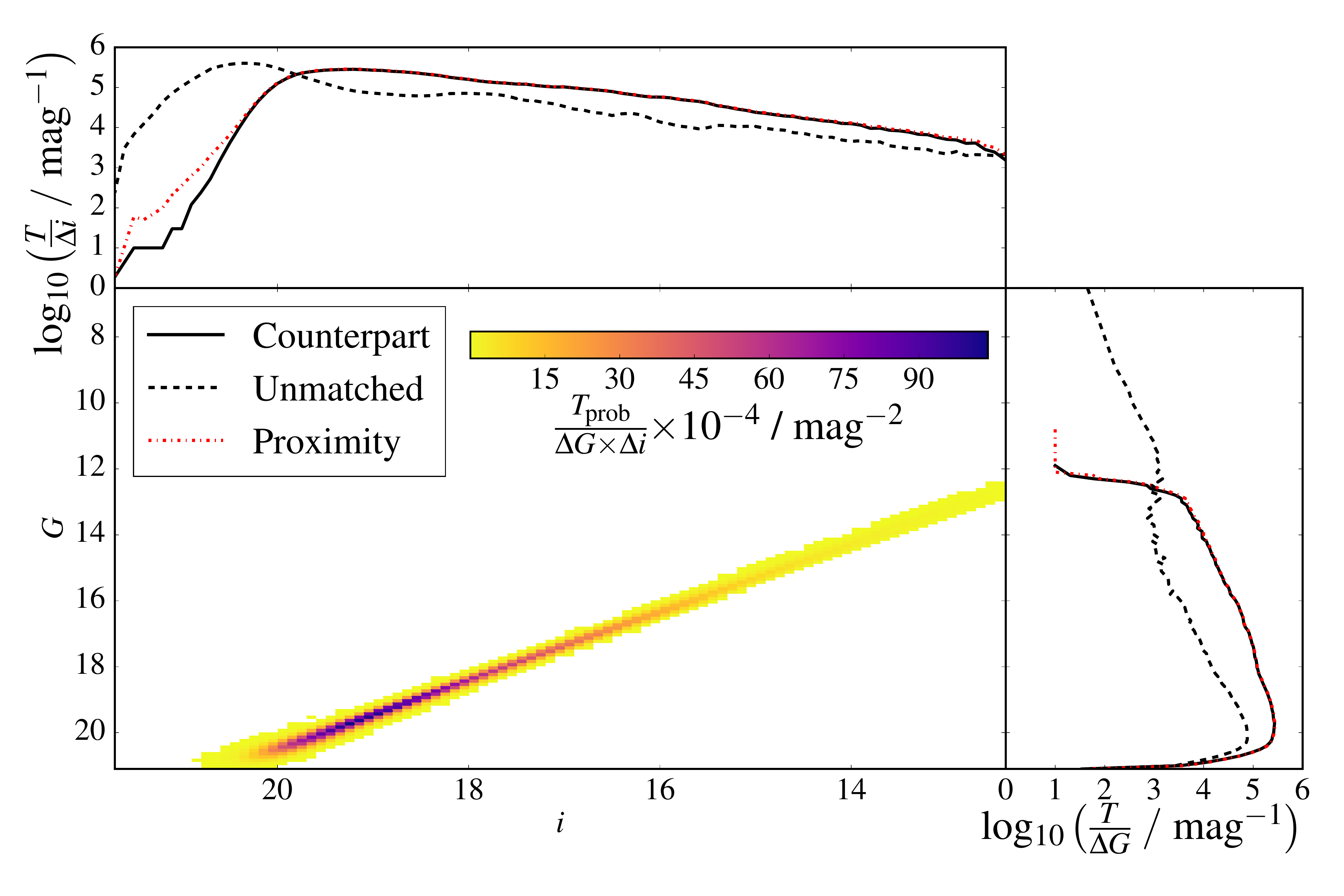}
    \caption{Figure showing the distributions for the probability matching of \textit{Gaia} and IPHAS in a 25 square degree region of the Galactic plane, in the $G$ and $i$ filters respectively. The middle panel shows a 2D histogram of probability-based counterparts in each small magnitude-magnitude bin. As expected from two similar optical passbands, the counterpart magnitude trend is roughly linear with decreasing brightness. The top and side panels show the number density of sources in each filter individually (i.e., the total number of stars returned as counterparts with a specific $G$ magnitude) in the solid black lines. Also shown in the inset figures are the unmatched star number densities (dotted black lines) and a 3" proximity-based match (red dot-dashed lines). The counterparts returned by proximity- and probability-based matches agree for most magnitude ranges. However, in the case of proximity matches we see an increase in the number of bright \textit{Gaia} counterparts that match to faint IPHAS objects, which the probability-based match rejects. Colourmap only displayed for those bins with densities $\geq500\mathrm{mag}^{-2}$.}
    \label{fig:iGprobvsproxmatchc}
\end{figure*}

We begin with the case of two optical catalogues, \textit{Gaia} and IPHAS. \textit{Gaia} saturates at a brighter magnitude than IPHAS, while IPHAS has a fainter completeness limit, which allows us to test our matching in the case of differing dynamical ranges. Figure \ref{fig:iGprobvsproxmatchc} shows the distributions of counterpart and unmatched stars for \textit{Gaia} $G$ and IPHAS $i$, comparing a 3" nearest-neighbour match to the probabilistic matching, accepting only those islands in which the most likely permutation is more probable than all other permutations. This proximity match is larger than our maximum island acceptance radius, resulting in a small number ($\lesssim 1\%$) of cases where we have a proximity match but no probability-based match based on the rejection of association during the island creation. However, these objects are rejected on both astrometric and photometric grounds, and we do not consider them further.

Several things need to be checked, using Figure \ref{fig:iGprobvsproxmatchc}, before we can be confident that the method correctly matches objects. First, stars in \textit{Gaia} that correspond to the saturated region in IPHAS should be returned as unmatched stars. The matched stars returned are shown as solid black lines in the side panels of Figure \ref{fig:iGprobvsproxmatchc}, and we can see a clear rejection of any match for stars of $G \lesssim 13$ (i.e., those detections saturated in IPHAS). Second, given the nature of matching two catalogues in the optical, we should return all stars as being matches in the dynamical range of the two catalogues. Comparing the matches in $13 \lesssim i \lesssim 19$, we can contrast our matches with a naive 3" proximity match, shown as the solid black lines and red dot-dashed lines in the side panels of Figure \ref{fig:iGprobvsproxmatchc} respectively. The probability-based matches return almost all of the proximity-based matches, as expected. Those unmatched objects in this region of overlapping dynamical ranges between the two catalogues are unexpected, with approximately one in five objects in either catalogue in this brightness range failing to return a match. However, over 80\% of these objects have no counterpart in the opposing catalogue within 5" (Section \ref{sec:catalogueapplication}), and are simply objects whose counterpart was rejected from the cleaned catalogues by our selection criteria (Table \ref{tab:flagtable}). The remaining 20\%, which do have a proximity match, are discussed later. Third, we wish to remove any potential mismatches between faint IPHAS objects and brighter \textit{Gaia} stars. Fainter than $i = 20$, we see a decrease in the number of counterparts returned by the probabilistic match, compared to the traditional proximity match (black solid lines vs red dashed lines in inset figures to Figure \ref{fig:iGprobvsproxmatchc}). One in four proximity matches is rejected as a probabilistic match fainter than $i \simeq 20$, a minority of which are systematically perturbed true matches and also discussed below. The loss rate increases by $i \simeq 21$ to four in every five proximity match pairs being assigned as unrelated, unmatched objects by the probability-based match. These rejections are mostly IPHAS objects too faint in $G$ to be detected, but serendipitously close to an unrelated bright \textit{Gaia} object, flagged in IPHAS. They have therefore been picked up as an unphysical match, and would be paired without the addition of the magnitude information.

\begin{figure}
    \centering
    \includegraphics[width=\columnwidth]{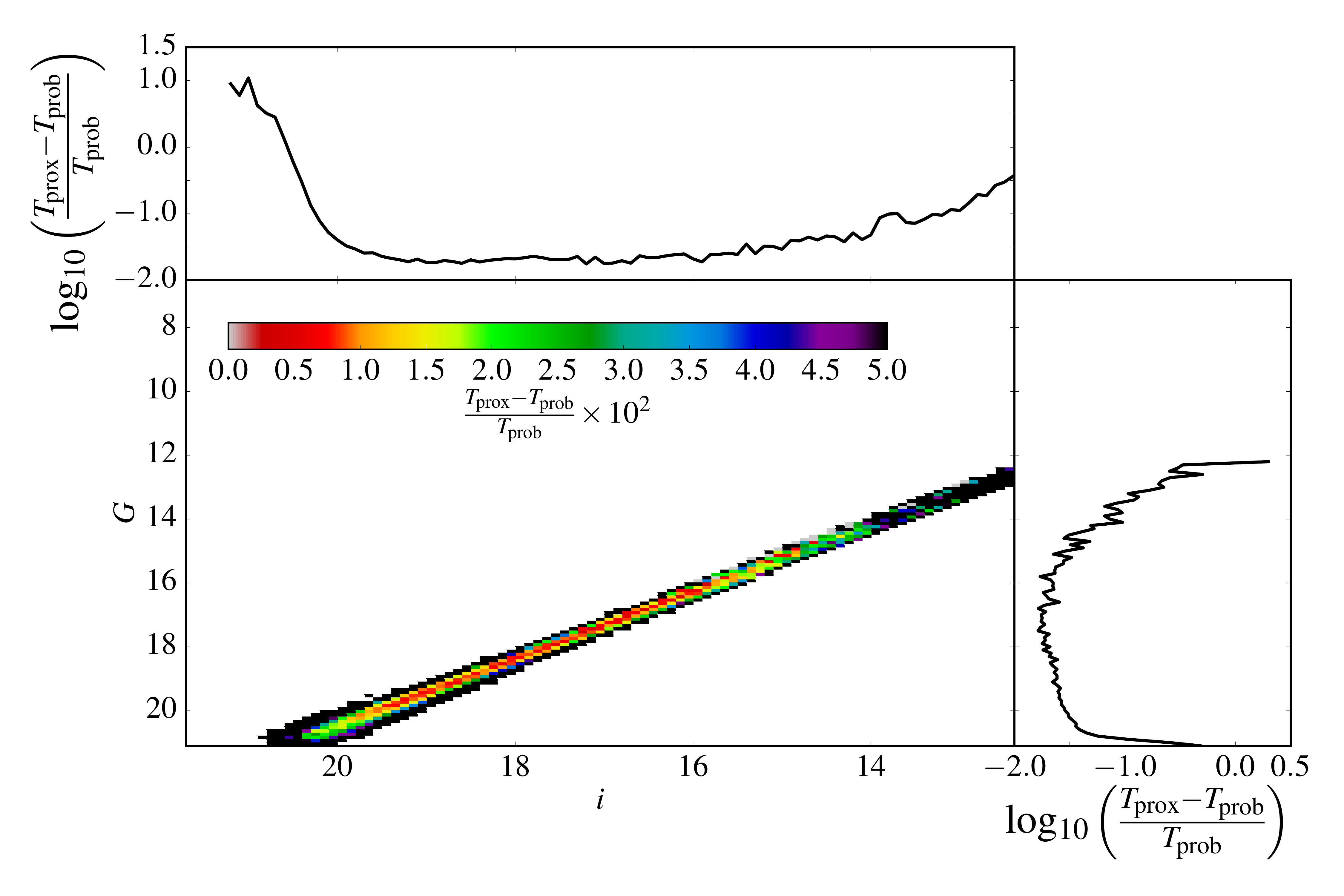}
    \caption{Figure showing the relative difference in number of objects returned for an IPHAS-\textit{Gaia} cross-match for 25 square degrees of the Galactic plane. Main panel shows the relative difference between the probability- and proximity-based matches for each small magnitude-magnitude bin, while the inset panels show the relative difference for each magnitude. At bright magnitudes a consistent rejection of matches occurs for $\lesssim3\%$ of objects. However, at fainter magnitudes ($i \gtrsim 20$) rejection of proximity matches occurs at a higher rate, caused in part by the assumption that the IPHAS AUF is purely Gaussian. The assumption of Gaussianity will cause the rejection of those objects in the non-Gaussian tails caused by systematic perturbations such as contamination due to faint, unresolved objects in the IPHAS PSF \citep{2017MNRAS.468.2517W}. Bins shown in main panel are the same as those which met the criterion in Figure \ref{fig:iGprobvsproxmatchc}.}
    \label{fig:iGmissing}
\end{figure}

We do return a small fraction of objects as field stars at brighter magnitudes that proximity matching assigns as counterparts, and should consider this population in more detail. Figure \ref{fig:iGmissing} shows the difference in the number density of probability- and 3" proximity-based matches. In the bright dynamic range of \textit{Gaia}, $12 \leq G \leq 17$, the typical loss of objects is $\simeq3\%$. However, this loss rate is across all IPHAS magnitudes, and includes $\lesssim1\%$ loss rate (i.e., one third of the total number of lost matches) of objects in the high counterpart density region of the magnitude-magnitude diagram. The rejections where the IPHAS magnitudes do not agree with the \textit{Gaia} brightness are reasonable and show the additional magnitude information correctly rejecting unlikely counterparts. However, the 1\% of rejections where the $i$ and $G$ magnitudes lie in the narrow range of accepted counterparts in both filters ought to be paired, and require further consideration.

When considering these unexpected rejections we can highlight the effect the magnitude information has on the counterpart matching scheme. However, before we are able to do so we must re-introduce the likelihood ratio (\citealp{Sutherland:1992aa}, \citealp{Fleuren:2012aa}, \citealp{Brusa:2005aa}, etc.), but split it into the photometric and astrometric components of, e.g., equation \ref{eq:rjnew3}. The photometric likelihood ratio, $\eta$, logarithmically balances the likelihood of matching magnitudes against the likelihood of the two stars being photometrically unmatched, given by

\begin{equation}
\eta \equiv \log_{10}\left(\frac{c(m_\gamma, m_\phi)}{f_\gamma(m_\gamma) f_\phi(m_\phi)}\right).
\label{eq:etaform}
\end{equation}
Equivalently, the astrometric likelihood ratio, $\xi$, is the logarithm of the comparison between the astrometric counterpart likelihood and the likelihood of the two objects being unrelated astrometrically, defined as

\begin{equation}
\xi \equiv \log_{10}\left(\frac{N_\mathrm{c}G}{N_\gamma N_\phi}\right).
\label{eq:xiform}
\end{equation}

\begin{figure}
    \centering
    \includegraphics[width=\columnwidth]{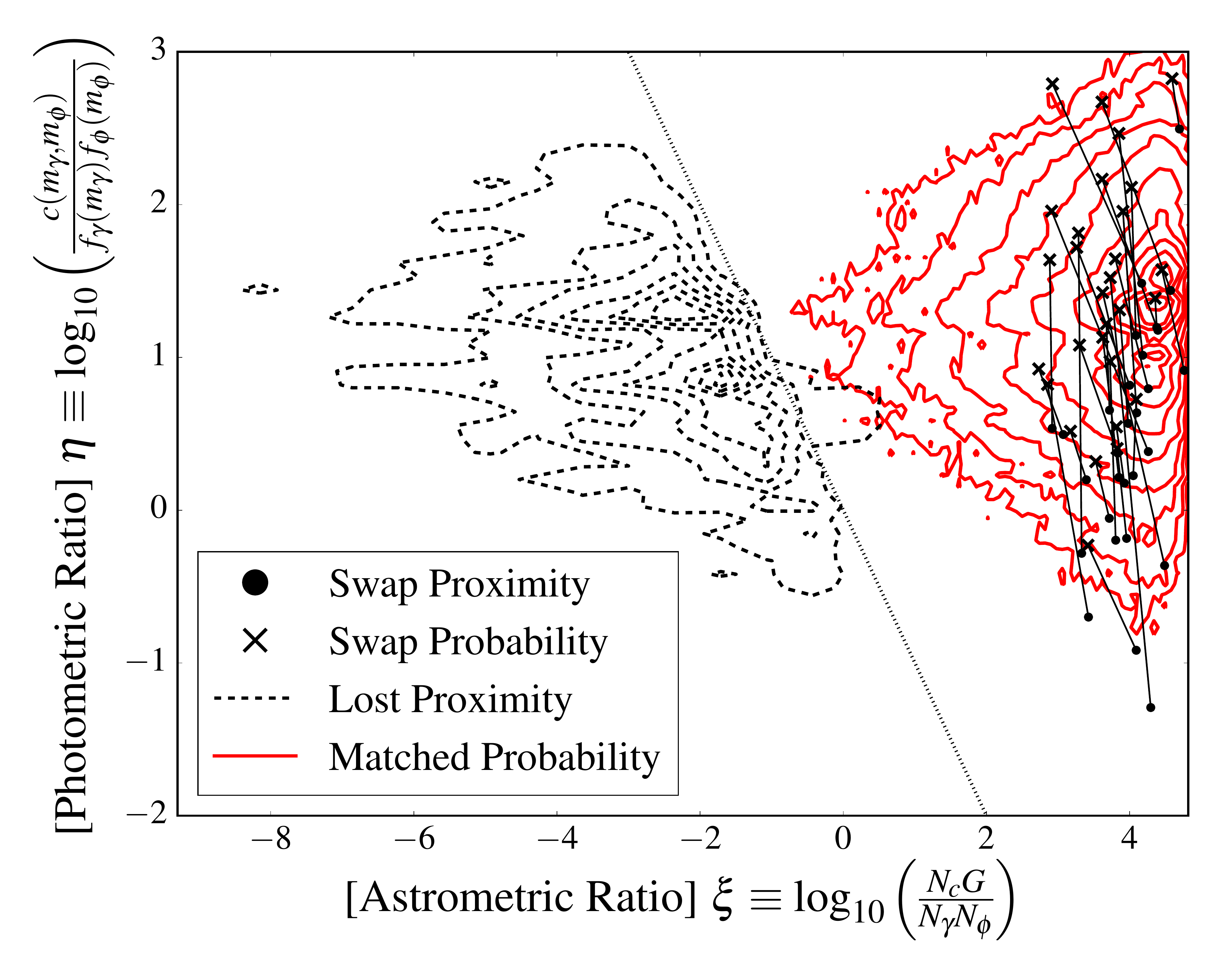}
    \caption{Figure showing the relative likelihoods of matched IPHAS and \textit{Gaia} stars, for a 25 square degree section of the Galactic plane. Here we are comparing the two likelihood ratios, photometric and astrometric, for the matches between the datasets. Red solid contours show the area of the plot occupied by the majority of the probability-based matches, while the black dashed contours show the area occupied by objects which were proximity matched to 3", but failed to return a probability-based match. Additionally, the connected lines are the cases where stars were proximity matched to one object, but returned a different probability-based match. These likelihood ratios are denoted by crosses for the probability-based match, circles for the proximity-based match, and are connected by a solid black line. Dotted line $\eta + \xi = 0$ represents a combined likelihood ratio of unity; equal chance between the two hypotheses.}
    \label{fig:iGlikeratiocomp}
\end{figure}

Consider Figure \ref{fig:iGlikeratiocomp}, which shows the main locus of those objects matched successfully by the probabilistic matching process (red solid contours). Also shown, in black dashed contours, is the area occupied in the ratio-ratio space by those objects that are returned by a proximity-based matching process but not by a probability-based match (i.e., those objects in Figure \ref{fig:iGmissing}). The vast majority of objects lost between the two processes are not lost due to low photometric chance. In fact, the contours lie in roughly the same region in $\eta$, but the lost objects have likelihood ratios six orders of magnitude lower in astrometry, compared to the main matched set. In both cases, the average improvement to the likelihood ratio that $\eta$ gives is approximately a 10-fold increase in probability. These high photometric likelihood but low astrometric likelihood objects are those whose astrometric positions are perturbed by systematic effects. They are perturbed to such a degree that they fall outside the maximum separation allowed by a Gaussian AUF \citep{2017MNRAS.468.2517W}. They are still within 3", however, and are therefore still picked up by a proximity match. This lowers their astrometric likelihood ratio until they become more likely unrelated objects than counterparts to the same source, as defined by the dotted line $\xi + \eta = 0$. We can distinguish these ``incorrect'' losses from truly rejected proximity matches by comparing both the photometric and astrometric likelihood ratios. While those matches that should not have been lost are only lost on astrometric grounds, a serendipitous proximity match has both poor photometric and astrometric likelihood ratios. We also see a few objects whose astrometric likelihood ratios are very high, but have photometric ratios slightly below one. These are the rare cases where objects coincidentally have magnitudes more typical of unrelated field objects (e.g., uncommon stellar types, non-stellar sources which have not been removed from during the data reduction process, etc.). However, their sky proximity is so overwhelmingly unlikely if they were unrelated that they simply must be detections of the same original object.

We can also consider the few cases in the set where one star has ``skipped'' over its closest neighbour and been matched with a nearby, but more distant, counterpart, similar to the example laid out in Section \ref{sec:matchqualitative}. In these cases the sky separation has increased, decreasing slightly our probability density $G$, but trading off against a large increase in photometric likelihood, as seen in Figure \ref{fig:iGlikeratiocomp} as the connected lines. This demonstrates the value of the additional information gained by using the photometry, allowing for the avoiding the pairing of two unrelated but serendipitously located objects.

\subsubsection{IPHAS vs 2MASS}
\label{sec:iphasvs2mass}
Next, we can compare the matches between IPHAS and 2MASS. For this matching process, however, we do not have a one-sided astrometric precision between our catalogues, because both IPHAS and 2MASS both have similar, $\simeq$0.05" positional precision in their bright, non-saturated regimes. This means that neither catalogue would be the obvious choice to map the other onto in an asymmetric matching fashion. It is therefore an important test of the symmetrisation of the photometric probabilities to the two-directional case.

\begin{figure*}
    \centering
    \includegraphics[width=\textwidth]{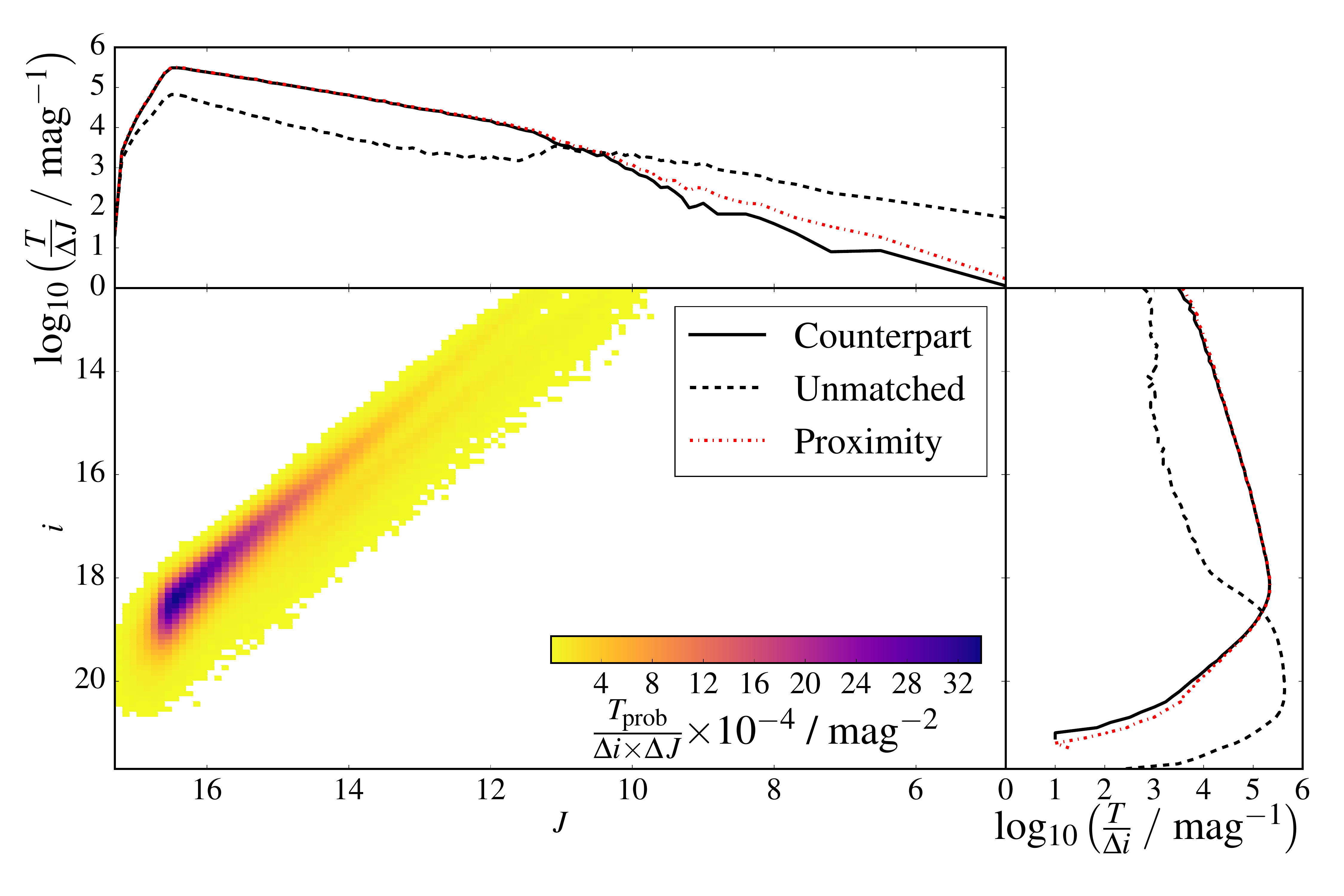}
    \caption{Figure showing the distributions of probability matched counterpart stars for 2MASS and IPHAS in a 25 square degree region of the Galactic plane, in the $J$ and $i$ filters respectively. Figure layout and colourbar are the same as Figure \ref{fig:iGprobvsproxmatchc}. Note the comparison to the proximity-based matches, where stars $J \leq 10$ are incorrectly assigned as matched to stars $i \geq 20$.}
    \label{fig:iJprobvsproxmatchc}
\end{figure*}

We successfully recover counterparts and unmatched stars, shown in Figure \ref{fig:iJprobvsproxmatchc}, in the correct magnitude ranges as with the IPHAS-\textit{Gaia} case above. Here we see a larger spread in accepted magnitudes in one catalogue for a given brightness in the other (i.e., a larger spread in $i-J$ colour). This is due mostly to the effects of differential extinction affecting the optical and near infra-red detections to differing degrees. Both IPHAS and now 2MASS contribute to the non-Gaussian tails in the wings of the AUFs. This means we still suffer from the rejection of several percent of likely counterparts at $i\simeq18$, in a similar effect to that described in Section \ref{sec:iphasvsgaia}. Additionally, we see an increase in the rejection of the pairing of faint IPHAS objects with bright 2MASS objects, as shown in the larger differences between the solid black and red dot-dashed lines in the side panels of Figure \ref{fig:iJprobvsproxmatchc}.

We see an effect which is not seen in the \textit{Gaia}-IPHAS case. In the case of the likelihood ratio comparison, we have some cases where both the astrometric and photometric likelihood ratios are increased by changing to a more distant counterpart, compared with that returned from proximity matching. These are the cases where a very faint object, which therefore has large astrometric uncertainties, is slightly closer to a bright object than another bright, and therefore astrometrically precise, object. This decrease in astrometric uncertainty leads to an increase in $G$, and thus $\xi$. The previously seen increase in $\eta$ is still observed, as the brighter object is correctly assigned as the counterpart.

\subsection{Summary}
\label{sec:applicationsummary}
In this section we applied the probability-based matching scheme to three test photometric catalogues, for the cases of \textit{Gaia} matched with IPHAS and IPHAS matched with 2MASS. We used the method as described in Sections \ref{sec:matchequation} through \ref{sec:makecandfcomp}. In both cases, we confirm the method correctly returns the majority of proximity-based matches. 

We discussed the key areas of the magnitude-magnitude space where the number of probabilistic matches deviates from the number of proximity matches. We conclude that the method is correctly rejecting some faint, proximity matched objects and assigning a brighter, more distant object as the counterpart. Additionally, we reject some proximity matches which are the proximity pairing of two different objects, matched accidentally. One object is lost (through, e.g., saturation or a poor detection) in catalogue $\gamma$ but within the dynamical range of catalogue $\phi$, while the other object is too faint to be included in catalogue $\phi$ but detected with good signal in catalogue $\gamma$. While we also reject some likely counterparts (i.e., two detections with similar magnitudes in similar passbands which we would expect to be the same source), we show these failed matches are lost based on their astrometry rather than their photometry. The assumption of pure Gaussian AUFs leads to unphysically small astrometric probabilities when objects are systematically perturbed to large separations relative to their astrometric uncertainties. The factor of approximately 10 increase in probability introduced with the addition of the photometric likelihoods is simply unable to overcome such low astrometric likelihood ratios.

In all cases, the additional parameter space from the magnitude information contributes to the resultant posterior probabilities. However, if the choice is made to model the probability density of star separations in detail, rather than using a simple cut-off radius, then it is critical that the AUFs are modelled properly. Correct AUF descriptions would minimise the rate of false non-pairings, allowing the photometric probabilities to distinguish between true and false matches.

\section{Extension to Multiple Catalogues}
\label{sec:multiplecatalogues}
\begin{figure}
    \centering
    \includegraphics[width=\columnwidth]{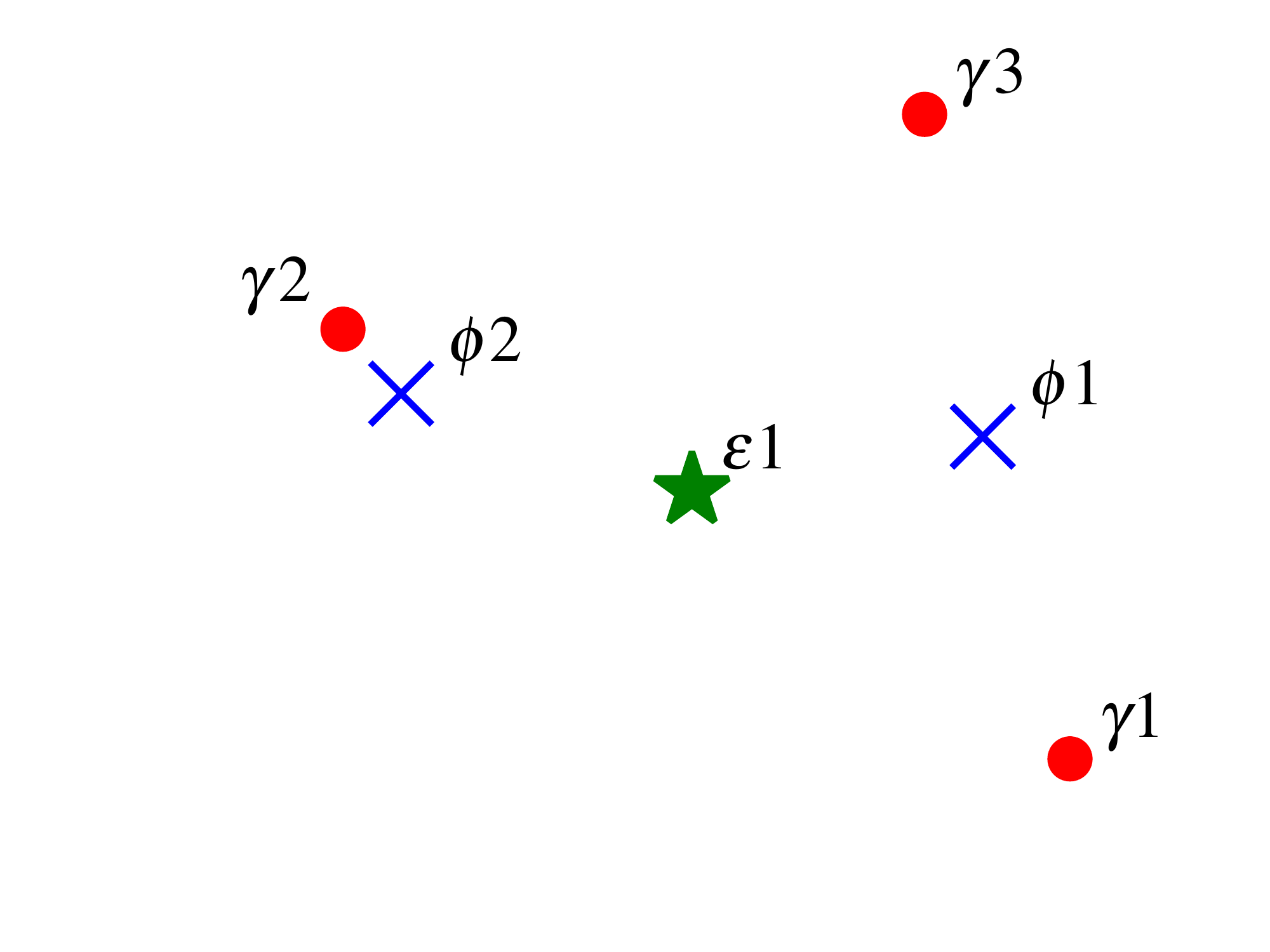}
    \caption{Figure showing an arrangement of potential matches from three theoretical catalogues. In this scenario one star is seen in all three catalogues as $\gamma$1, $\phi$1, and $\epsilon$1 respectively; $\gamma$2 and $\phi$2 are the same star recorded in two catalogues; and a third star, $\gamma$3, is only seen in one catalogue. Catalogue $\gamma$ sources are denoted by red circles, catalogue $\phi$ sources are shown as blue crosses, and the single green star source is from catalogue $\epsilon$.}
    \label{fig:multiplecatalogues}
\end{figure}

So far, in Sections \ref{sec:matchqualitative} to \ref{sec:catalogueapplication}, we have only considered the case where we are matching one catalogue against another. However, oftentimes we wish to match multiple catalogues to extend our wavelength coverage. Imagine a hypothetical scenario for a three-catalogue match. Shown in the schematic in Figure \ref{fig:multiplecatalogues} are three example stars, observed in three example catalogues. Catalogue $\gamma$ observed three stars in the small field of view in consideration, denoted $\gamma1$, $\gamma2$, and $\gamma3$, shown as red circles. Catalogue $\phi$, shown as blue crosses, observed two of the stars: $\phi1$ and $\phi2$. Finally, our third catalogue $\epsilon$ only recorded a measurement for $\epsilon1$, shown in Figure \ref{fig:multiplecatalogues} as the green star. 

We could potentially iterate all possible permutations of this set, asking what the probability is that, e.g., stars $\gamma$2 and $\phi$2 are counterparts to each other, star $\gamma$3 is uncorrelated and stars $\gamma$1, $\phi$1, and $\epsilon$1 are all counterparts of the same object. Considering all possibilities would require extensions to $c$, asking what the likelihood of counterparts having magnitudes $m_{\gamma1}$, $m_{\phi1}$, and $m_{\epsilon1}$ was, as well as an extension to $G$, given now as 

\begin{align}
\begin{split}
G{'}(\Delta x_{\gamma1\phi1}&, \Delta y_{\gamma1\phi1}, \Delta x_{\gamma1\epsilon1}, \Delta y_{\gamma1\epsilon1}) = \\
\int\limits_{-\infty}^{+\infty}\!\int\limits_{-\infty}^{+\infty}\!&h_\gamma(x_{\gamma1} - x_0, y_{\gamma1} - y_0)h_\phi(x_{\phi1} - x_0, y_{\phi1} - y_0)\,\times\\&h_\epsilon(x_{\epsilon1} - x_0, y_{\epsilon1} - y_0)\,\mathrm{d}x_0\,\mathrm{d}y_0,
\label{eq:Gextension}
\end{split}
\end{align}
where $h_\gamma$, $h_\phi$ and $h_\epsilon$ are the astrometric distributions of the three catalogue respectively.

However, the complexity of the problem increases geometrically, and it quickly becomes impractical to treat even three catalogues simultaneously. In cases where more than two catalogues are required, sequential matching, starting from the two most astrometrically precise catalogues and working towards the least precise astrometry, is recommended. Starting with a match between catalogues $\gamma$ and $\phi$, we create catalogue $\gamma\phi$, which contains matches between both catalogues, unmatched catalogue $\gamma$ objects and unmatched catalogue $\phi$ objects. Subsequently we take catalogue $\gamma\phi$ and match it with catalogue $\epsilon$, creating a catalogue which contains matches between $\gamma$, $\phi$, and $\epsilon$; $\gamma$ and $\epsilon$ matches; $\phi$ and $\epsilon$ matches; $\gamma$ and $\phi$ matches; and objects in catalogues $\gamma$, $\phi$, and $\epsilon$ which do not match to either of the other two catalogues.

For example, we might require a composite catalogue with optical detections (e.g., IPHAS), near-IR sources (e.g., 2MASS), and detections at longer wavelengths (e.g., \textit{Spitzer}; \citealp{2004ApJS..154....1W}). In this instance we might first match IPHAS and 2MASS (see Section \ref{sec:iphasvs2mass}), creating our first sequential composite cross-match catalogue, and then match \textit{Spitzer} with this new catalogue. However, when matching the second time, we have removed from our normalisations, such as those in equation \ref{eq:hypoA}, any hypothesis where any sources paired during the IPHAS-2MASS match are not paired. However, we can choose to only accept high probability classifications from previous iterations of the sequential matching (see Section \ref{sec:reducecompcomplex} for more details). These relatively certain classifications will have low probabilities of any other hypothesis, and the exclusion of the hypothesis of previous IPHAS-2MASS matches being unrelated will have little impact on the conclusions drawn. Thus we can reduce the complexity of a multi-catalogue cross-match into several two catalogue cross-matches.

The only concession we have to make is in the careful treatment of equation \ref{eq:Gextension} (cf. equation \ref{eq:baymatch5c} for the original two-catalogue case). We cannot easily split equation \ref{eq:Gextension} into sequential terms, and in order to do so we have to ``update'' the position of a counterpart pair merge after each cross-match, which is why it is recommended that the most precise catalogues are used initially. We then use the weighted mean position of the two matched stars as the new position. Updating the position of the source in this way is comparable to section 5.1 of \citet{Pineau:2017aa}, although since one cannot guarantee Gaussianity of the distributions (see \citealp{2017MNRAS.468.2517W}) this becomes

\begin{equation}
x_\mathrm{new} = \frac{\iint\limits_{-\infty}^{+\infty}\!h_\gamma(x_{\gamma1}\! -\! x_0, y_{\gamma1}\! -\! y_0)h_\phi(x_{\phi1}\! -\! x_0, y_{\phi1}\! -\! y_0)x_0\,\mathrm{d}x_0\,\mathrm{d}y_0}{(h_\gamma*h_\phi)(x_{\gamma2}-x_{\phi2}, y_{\gamma2}-y_{\phi2})}
\label{eq:meanposition}
\end{equation}
with analogous arguments for $y_\mathrm{new}$. While we can relatively easily update the position of the star in our new cross-matched catalogue, it is less straightforward to handle the updated AUF. We therefore recommend simply using the appropriate covariance matrix and AUF of the most positionally precise of the two merged stars. 

In the era of increasingly precise datasets, such as \textit{Gaia}, the complication of sequential matching becomes increasingly negligible, as equation \ref{eq:Gextension} simply returns

\begin{align}
\begin{split}
G{'}(&\Delta x_{\gamma1\phi1}, \Delta y_{\gamma1\phi1}, \Delta x_{\gamma1\epsilon1}, \Delta y_{\gamma1\epsilon1}) \\&= h_\phi(x_{\phi1} - x_{\gamma1}, y_{\phi1} - y_{\gamma1})h_\epsilon(x_{\epsilon1} - x_{\gamma1}, y_{\epsilon1} - y_{\gamma1})
\label{eq:Gextensionreduce}
\end{split}
\end{align}
in the limit of $h_\gamma(x_{\gamma1} - x_0)\to\delta(x_{\gamma1} - x_0)$. Effectively, we only have to ask the probability of our two other catalogues being drawn from the order-of-magnitude more precise third position.

\section{Conclusions}
\label{sec:conclusions}
We have developed a new symmetric method for assigning stars between two catalogues as either counterparts, or unrelated and unmatched stars. We use the extra information gained from the measured photometric magnitudes of the stars to more accurately accept or reject star pairings. Our more general formalism for the astrometric probability formally describes the handling of astrometric uncertainties in an equal fashion. It also allows for a more general inclusion of systematic astrometric effects such as proper motion or contamination caused by stellar crowding. We have also expanded the treatment of photometric probabilities to a two-directional treatment, asking the probability of a star having the detected magnitudes of both objects. This new method also allows for the possibility of multiple choices of counterpart for stars in each catalogue. Additionally, we showed how to extend the method to multiple catalogues. 

We tested the method on three catalogues: IPHAS, 2MASS, and \textit{Gaia}. We showed that the method correctly returns counterparts in the expected regimes of shared dynamical range between two given catalogues. When compared to a 3" proximity-based match, we successfully return more unassigned, unmatched objects at very bright and very faint magnitudes, outside of the dynamical range of the opposing catalogue. We also show that the method works when applied to two catalogues of similar astrometric precision, with a truly symmetric handling of the assigning of counterparts between catalogues. In all catalogue match cases, and in all brightness regimes, the inclusion of the photometric likelihoods allowed for a more robust determination of the corresponding objects between catalogues, providing on average a factor 10 improvement to the Bayes' factor. This provides the ability to break nearest-neighbour and pure astrometric probability match degeneracies.

The nature of the method gives the flexibility to choose a probability above which to accept counterparts, allowing for the option of only selecting very likely joins between catalogues, giving the confidence in the resulting SEDs.

\section*{Acknowledgements}
\label{sec:acknowledge}
The authors thank the referee for their thorough report and useful comments, which helped us to improve this paper. TJW acknowledges support from an STFC Studentship. TN is funded by a Leverhulme Trust Research Project Grant. This work has made use of the SciPy \citep{scipy}, NumPy \citep{numpy}, Matplotlib \citep{matplotlib}, and F2PY \citep{f2py} Python modules, and NASA's Astrophysics Data System.

This paper makes use of data obtained as part of the INT Photometric H$\alpha$ Survey of the Northern Galactic Plane (IPHAS, www.iphas.org) carried out at the Isaac Newton Telescope (INT). The INT is operated on the island of La Palma by the Isaac Newton Group in the Spanish Observatorio del Roque de los Muchachos of the Instituto de Astrofisica de Canarias. All IPHAS data are processed by the Cambridge Astronomical Survey Unit, at the Institute of Astronomy in Cambridge. The bandmerged DR2 catalogue was assembled at the Centre for Astrophysics Research, University of Hertfordshire, supported by STFC grant ST/J001333/1. 

This publication makes use of data products from the Two Micron All Sky Survey, which is a joint project of the University of Massachusetts and the Infrared Processing and Analysis Center/California Institute of Technology, funded by the National Aeronautics and Space Administration and the National Science Foundation. 

This work has made use of data from the European Space Agency (ESA) mission \textit{Gaia} (\url{http://www.cosmos.esa.int/gaia}), processed by the \textit{Gaia} Data Processing and Analysis Consortium (DPAC, \url{http://www.cosmos.esa.int/web/gaia/dpac/consortium}). Funding for the DPAC has been provided by national institutions, in particular the institutions participating in the \textit{Gaia} Multilateral Agreement. 

\appendix
\section{Reduction to One-Sided Case}
\label{sec:matchreduction}
We have presented a symmetric approach to the probability-based matching procedure treated asymmetrically by several previous authors (e.g., \citealp{Sutherland:1992aa}, \citealp{Naylor:2013aa}, \citealp{Rutledge:2000aa}). To verify the validity of the formalism, we must check that the equations reduce to the one-sided set of equations in the correct limits. As our formalism is based upon that of \citet{Naylor:2013aa}, we shall confirm that we can recover their equations in this section.

The differences introduced in equations \ref{eq:baymatch11} and \ref{eq:baymatch12}, compared with equations 6 and 7 of \citet{Naylor:2013aa}, come from the reduced dimensionality of the problem, as well as several underlying assumptions. If it were possible to treat, e.g., X-ray sources as independent entities, each with a unique set of potential counterparts, we could break the larger catalogue up into smaller ones, each of which only containing one source, resulting in an effective catalogue length of one. This is equivalent to assuming the catalogue has internal independency. In this case, it is obvious that the number of matches is either zero or one. One can then split equation \ref{eq:baymatch12} into two cases. First, the case where $M = 0$, with one permutation allowed in $\zeta$ and $\lambda$. Second, the case of $M = 1$, where $\lambda$ still only has one permutation, due to its catalogue having length one. This reduces the triple sum to a sum over $\gamma$, each star being the counterpart in turn, which is the sum over $j$ in equations 6 and 7 of \citet{Naylor:2013aa}. Equivalently, starting from equations \ref{eq:rjnew3} and \ref{eq:rjnew4} we can recover equations 6 and 7 of \citet{Naylor:2013aa} by forcing the number of elements over which $i$ is summed to be one, which removes the $i\neq s$ sum, and reduces the sum over $s$ and $t$ to just one over $t$.

This reduction in dimensionality is possible if and only if the separation between catalogue $\phi$'s stars is much greater than the average radial offset of their counterparts in catalogue $\gamma$. This means there is no overlap and no two catalogue $\phi$ stars can possibly have the same star in catalogue $\gamma$ within a given radial offset of both stars. Additionally, \citet{Naylor:2013aa} made the assumption that the two catalogues' magnitudes are independent of each other, and thus $c(m_k, m_l) = c(m_k)\,c(m_l)$. Finally, two implicit assumptions were made. The first is that $c(m_l) = f_\phi(m_l)$. Second, the assumption was made that catalogue $\phi$ is complete, meaning that we do not require the symmetrisation of the counterpart magnitude probability density in Section \ref{sec:matchfunctions}, effectively setting $p_\phi = 1$. 

To introduce the concept of $X$ into our equations (see Table 1 of \citealp{Naylor:2013aa}) we define it as the fraction of stars with counterparts in catalogue $\phi$,

\begin{equation}
X = \frac{N_\mathrm{c}}{N_\phi + N_\mathrm{c}}.
\label{eq:baymatch5j}
\end{equation}
Rearranging the terms, we obtain

\begin{align}
\begin{split}
\frac{N_\mathrm{c}}{N_\phi} = \frac{X}{1-X}.
\label{eq:baymatch5k}
\end{split}
\end{align}

We can now reproduce the correct ratios found in $p(H_a|D)$. To do so, we start with our original equations \ref{eq:rjnew3} and \ref{eq:rjnew4}, restated in their compact notation (Section \ref{sec:onematchform}) as

\begin{align}
\begin{split}
&P(H_a|D) = \\
&\frac{N_\mathrm{c}G^{kl}_{\gamma\phi}c^{kl}_{\gamma\phi}\prod\limits_{i \neq k}\,N_\gamma f_\gamma^i\prod\limits_{j \neq l}N_\phi f_\phi^j}{\prod\limits_i\,N_\gamma f_\gamma^i\prod\limits_jN_\phi f_\phi^j + \sum\limits_s\,\sum\limits_t\,N_\mathrm{c}G^{st}_{\gamma\phi}c^{st}_{\gamma\phi}\prod\limits_{i \neq s}\,N_\gamma f_\gamma^i\prod\limits_{j \neq t}N_\phi f_\phi^j},
\label{eq:rjnew3b}
\end{split}
\end{align}
and
\begin{align}
\begin{split}
&P(H_0|D) = \\
&\frac{\prod\limits_i\,N_\gamma f_\gamma^i\prod\limits_jN_\phi f_\phi^j}{\prod\limits_i\,N_\gamma f_\gamma^i\prod\limits_jN_\phi f_\phi^j + \sum\limits_s\,\sum\limits_t\,N_\mathrm{c}G^{st}_{\gamma\phi}c^{st}_{\gamma\phi}\prod\limits_{i \neq s}\,N_\gamma f_\gamma^i\prod\limits_{j \neq t}N_\phi f_\phi^j}.
\label{eq:rjnew4b}
\end{split}
\end{align}
First we set the length of catalogue $\gamma$ to one, which removes the product $\prod\limits_{i \neq k}\,N_\gamma f_\gamma^i$ and reduces the product $\prod\limits_i\,N_\gamma f_\gamma^i$ to $N_\gamma f_\gamma^k$. We also multiply and divide any terms containing $\prod\limits_{j \neq l}N_\phi f_\phi^j$ in equations \ref{eq:rjnew3b} and \ref{eq:rjnew4b} by $N_\phi f_\phi^l$, for both $l$ and $t$.

Switching back to our full notation, we then divide all terms in equations \ref{eq:rjnew3b} and \ref{eq:rjnew4b} by

\begin{equation}
N_\gamma f_\gamma(m_k)\prod\limits_jN_\phi f_\phi(m_j).
\label{eq:thingwecancancel}
\end{equation}
This gives us

\begin{align}
\begin{split}
P(H_a|D) = \frac{\frac{N_\mathrm{c}G(\Delta x_{kl}, \Delta y_{kl})c(m_k, m_l)}{N_\gamma f_\gamma(m_k)N_\phi f_\phi(m_l)}}{1 + \sum\limits_l\,\frac{N_\mathrm{c}G(\Delta x_{kl}, \Delta y_{kl})c(m_k, m_l)}{N_\gamma f_\gamma(m_k)N_\phi f_\phi(m_l)}}
\label{eq:baymatchsimplepha}
\end{split}
\end{align}
and

\begin{align}
\begin{split}
P(H_0|D) = \frac{1}{1 + \sum\limits_l\,\frac{N_\mathrm{c}G(\Delta x_{kl}, \Delta y_{kl})c(m_k, m_l)}{N_\gamma f_\gamma(m_k)N_\phi f_\phi(m_l)}},
\label{eq:baymatchsimpleph0}
\end{split}
\end{align}
re-introducing the likelihood ratio to our probabilities.

Therefore, after splitting $c(m_k, m_l)$ into $c(m_k)c(m_l)$; cancelling $c(m_l)$ and $f_\phi(m_l)$, assumed to be equivalent; substituting for equation \ref{eq:baymatch5k}; and multiplying by $1 - X$, we recover equations 6 and 7 of \citet{Naylor:2013aa},

\begin{align}
\begin{split}
P(H_a|D) = \frac{\frac{X g(\Delta x, \Delta y)}{N_\gamma}\frac{c(m_a)}{f_\gamma(m_a)}}{1 - X + \sum\limits_\alpha\frac{X g(\Delta x, \Delta y)}{N_\gamma}\frac{c(m_\alpha)}{f_\gamma(m_\alpha)}}
\label{eq:baymatch6}
\end{split}
\end{align}
and

\begin{align}
\begin{split}
P(H_0|D) = \frac{1 - X}{1 - X + \sum\limits_\alpha\frac{X g(\Delta x, \Delta y)}{N_\gamma}\frac{c(m_\alpha)}{f_\gamma(m_\alpha)}}.
\label{eq:baymatch7}
\end{split}
\end{align}
Note that the $g$ term of \citet{Naylor:2013aa} is our $G$, as they add a systematic uncertainty to their X-ray uncertainties, believed to reflect the infrared uncertainties, and thus it is a convolution of two Gaussians.

While the appendix derivation of \citet{Naylor:2013aa} required $P(H_0) = 1-X$ and $P\left(\widetilde{H}_0\right) = X$, our new derivation contains these implicitly as the ratio of counterparts per unit area to unmatched stars per unit area. We therefore have indifferent priors, assuming a flat prior across all hypotheses. This is required in our formalism due to the extension to a symmetric handling of stars in both catalogues, as well as the extension to multiple potential counterparts in each catalogue. The number densities of matched and unmatched objects can only be considered as simple Bayesian priors in the case where the information of only one catalogue is used, for one potential counterpart. However, the end result is identical, and the equations correctly reduce to their original forms in various limits.

%%%%%%%%%%%%%%%%%%%%%%%%%%%%%%%%%%%%%%%%%%%%%%%%%%

%%%%%%%%%%%%%%%%%%%% REFERENCES %%%%%%%%%%%%%%%%%%

% The best way to enter references is to use BibTeX:

\bibliographystyle{mnras}
\bibliography{../../PostgradPapers_jr2.bib} % if your bibtex file is called example.bib

%%%%%%%%%%%%%%%%%%%%%%%%%%%%%%%%%%%%%%%%%%%%%%%%%%

% Don't change these lines
\bsp	% typesetting comment
\label{lastpage}
\end{document}